# Electrocatalytic removal of persistent organic contaminants at molybdenum doped manganese oxide coated TiO$_2$ nanotube-based anode

*Natalia Sergienko[a,b], Elisabeth Cuervo Lumbaque[a,b], Nick Duinslaeger[a,b], Jelena Radjenovic[a,c*]*

*Catalan Institute for Water Research (ICRA), Scientific and Technological Park of the University of Girona, 17003 Girona, Spain*

*[b] University of Girona, Girona, Spain*

*[c] Catalan Institution for Research and Advanced Studies (ICREA), Passeig Lluís Companys 23, 08010 Barcelona, Spain*

*\* Corresponding author:*

*Jelena Radjenovic, Catalan Institute for Water Research (ICRA), c/Emili Grahit, 101, 17003 Girona, Spain*

Phone: + 34 972 18 33 80; Fax: +34 972 18 32 48; E-mail: jradjenovic@icra.cat

# Abstract

Electrooxidation is an attractive technique that can be effectively applied for the treatment of persistent organic contaminants, but its implementation in practice is limited by the lack of efficient and low-cost anode material that would not lead to the generation of chlorinated by products. Herein, we developed a novel anode based on $TiO_2$ nanotube array (NTA) coated with Mo-doped $Mn_xO_y$, and applied it for electrooxidation of persistent organic contaminants. The Ti/$TiO_2$ NTA–$Mn_xO_y$-Mo anode outperformed the commercial Ti/$IrO_x$-Pt anode and effectively oxidized organic contaminants, with a significantly reduced electric energy per order, and avoiding chlorine evolution reaction. Mo doping played a key role in the excellent performance of the synthesized anode as it favored the formation of oxygen vacancies in the host lattice and Mn and Mo species redox couples, which increased the oxidizing power of the anode and ensured its complete stability.

# 1. Introduction

Over the next few decades, climate change, aquifer depletion and population growth will greatly limit access to reliable and safe water sources [1]. Centralized water and wastewater treatment cannot cope well with the soaring water stress [2]. The well-established water treatment technologies fail to effectively remove new classes of organic contaminants, which include biocides, pharmaceuticals, and personal care products. Decentralization of the water treatment coupled with the introduction of new treatment methods can enable at-source, and point-of-entry or point-of-use removal of emerging organic contaminants and help alleviate the water crisis.

Electrochemical water treatment is a process with great potential to evolve into a platform technology for decentralized (waste)water treatment. Electrochemical systems are versatile and robust, they can operate at ambient temperature and pressure, and without any addition of chemicals [3]. They are modular, easily coupled with other treatment techniques into hybrid processes, and have small footprint, which is a crucial for a decentralized water treatment technology. Furthermore, electrochemical systems can be easily coupled with renewable energy sources such as photovoltaic solar cells, and thus represent a very attractive technology to meet the demands of the water-energy nexus [4].

The choice of the electrode material is extremely important in the electrochemical cell design. Anode materials typically used in water treatment include boron doped diamond (BDD), mixed metal oxide (MMO) and Magnéli phase electrodes. However, there are still several important limitations to overcome. First, commercial electrode materials display high activity for electro-oxidation of chloride to free chlorine [5], which reacts with the organic matter and forms highly toxic and persistent chlorinated organic byproducts [6, 7]. Moreover, high current densities typically required for the generation of strong oxidants (e.g., hydroxyl radicals, OH•) oxidize chloride to toxic and persistent chlorate and//or perchlorate anions. Considering that chloride is ubiquitously present in all natural waters, the avoidance of byproduct formation represents a major challenge in electrochemical wastewater treatment. Finally, the price of most of the commercially available anodes, e.g., ~€4200 per $m^2$ € (BDD) and 3000 € (MMO) per $m^2$ of anode surface area, limits the employed size of the electrode per volume/flow of treated water, resulting in low space-time yields of anodic oxidation [8, 9].

Hence, there is an urgent need for a low cost and environmentally safe anode material, that could effectively oxidize persistent organic contaminants, while avoiding the generation of halogenated byproducts. One of the possible candidates are materials based on manganese, a transitional metal characterized by a unique chemistry and exceptionally high catalytic activity. Manganese oxide can exist in a variety of oxidation states ranging from Mn II to Mn IV, and easily transition between them. This ability coupled with the high standard redox potential (e.g., 1.23 V for Mn IV) enables the involvement of manganese oxide into various redox reactions [10]. For instance, manganese oxides can oxidize organic contaminants often present in wastewater (e.g., substituted phenols and anilines) [11]. Hence, excellent catalytic activity of manganese, its low cost and earth abundance, as well as its relatively low toxicity, make manganese oxide-based anodes extremely attractive for environmental applications [10, 12].

Optimization and tailoring of manganese oxide-based materials towards specific applications became a subject of great interest in recent research. Several studies demonstrated that substitutional cationic dopants (e.g., $Ba^{2+}$, $Sn^{2+}$, $Co^{3+}$, $Ni^{2+}$, $Ag^+$, $Ce^{3+}$, $Mo^{2+}$, $V^+$, $Nb^+$ etc.) can significantly improve the catalytic activity and stability of manganese oxides through weakening of the bond between the metal lattice and oxygen atoms, and introduction of oxygen vacancies [13, 14]. Also, the incorporation of the cationic dopants enhances the electrical conductivity of manganese oxide, which is highly desirable for electrochemical applications [15].

In our previous work, we demonstrated the key role of substrate in the overall performance manganese oxide anode. Ti substrate with $TiO_2$ nanotube array (NTA) layer significantly enhanced the activity and stability of the resulting anode materials, especially in comparison with the carbon-based support often used for manganese oxide deposition [16]. In a recent study, manganese oxide-coated Ti/$TiO_2$ NTA was employed for the electrooxidation of phenol [17]. Nevertheless, manganese oxide-based materials are known to undergo reductive dissolution in the presence of organics, and new strategies are required to stabilize the active coating [18].

In this study, we synthesized a manganese oxide anode on a $TiO_2$ NTA via electrodeposition pathway and investigated the impact of molybdenum doping on its stability and electrochemical performance. Several material characterization techniques were employed to study the nature of the molybdenum incorporation into the electrodeposited manganese oxide and to elucidate the enhancing effect of Mo doping on the electrode stability, electrocatalytic activity and electrochemically active surface area. The resulting material was then applied for electrocatalytic oxidation of a set of model organic contaminants with diverse polarity and charge, and known to be persistent to chemical and electrochemical treatment, including antiviral medication acyclovir (ACY), antiepileptic drug carbamazepine (CBZ), nonsteroidal anti-inflammatory drug diclofenac (DCL), anti-bacterial agent triclosan (TCL) and analgesic tramadol (TMD) [19]. We investigated the influence of operating parameters such as applied potential and presence of anions (i.e., $NO_3^-$, $Cl^-$, $PO_4^{3-}$) on the electrocatalytic oxidation of the target contaminants. As a result of low removal efficiencies of persistent organic contaminants from sewage at conventional treatment plants, these contaminants are often encountered in drinking water [20, 21]. Hence, we evaluated the performance of the developed TiO$_2$ NTA–Mn$_x$O$_y$-Mo electrodes for their removal from real tap water. To gain more insight into the oxidation mechanisms of organic pollutants at the Mn$_x$O$_y$-Mo based anode, we identified major transformation products of CBZ oxidation and proposed plausible degradation pathways. In this study, we demonstrated for the first time that TiO$_2$ NTA coated with the catalytically active Mo-doped Mn$_x$O$_y$ can oxidize a variety of persistent organic contaminants at low anode potentials even in low conductivity solutions. The electrochemical treatment of water with low conductivity usually requires a significant amount of energy due to the high ohmic losses [6]. The ability to treat low conductivity solutions at low potentials can significantly reduce the energy requirements for the process, thus making electrochemical water treatment more energy and cost effective. Finally, the use of TiO$_2$ NTA–Mn$_x$O$_y$-Mo electrodes can effectively prevent the undesirable electrolysis of chloride into free chlorine, chlorate, and perchlorate. This is a significant benefit, as it ensures that the electrochemical treatment process remains safe and environmentally friendly, without generating harmful by-products.

## 2. Materials and methods

### 2.1. TiO$_2$ NTA - Mn$_x$O$_y$ and TiO$_2$ NTA - Mn$_x$O$_y$ Mo material synthesis

The pretreatment of 6 × 4 cm Ti mesh (not coated, 99.6% purity, De Nora, Italy) consisted of several steps. First, the meshes were polished by sandblasting to ensure complete removal of surface TiO$_2$ layer. Sandblasted Ti meshes were then cleaned thoroughly by sonication in the mixture of isopropanol, acetone, and methanol until the particles produced by the sandblasting were completely removed. The cleaned meshes were stored in ethanol solution to avoid gradual formation of the surface TiO$_2$ layer. Prior to the anodization, the mesh was dried in a nitrogen stream and subjected to 17% w/w HCl aqueous solution (Scharlab, Spain) at 75°C for 15 min to increase the surface roughness. Next, the Ti mesh was immersed in an electrolyte comprised of glycerol and deionized water (50:50 vol.%) with 0.5 wt.% $NH_4F$. The electrolyte was constantly stirred with a magnetic stirrer at 200 rpm. The polarization of the mesh was performed in a two-electrode setup vs a stainless steel split counter electrode. The potential of the cell was controlled with Autolab 302N potentiostat/galvanostat equipped with the voltage amplifier (Metrohm Autolab B.V., The Netherlands). The procedure included gradual increase of the potential from the open circuit (OC) to 20 V with the step of 0.2 V $sec^{-1}$, followed by the anodization at constant potential of 20 V during 2 hours. After the anodization, the $TiO_2$ NTA samples were soaked in deionized water for 12 hours and thermally treated in argon atmosphere at 400°C for 2 hours using a tubular oven (Nabertherm, Germany).

The electrodeposition of the $Mn_xO_y$ and Mo-doped $Mn_xO_y$ onto the synthesized $Ti/TiO_2$ NTA support was performed in a three-electrode setup at ambient temperature, using stainless steel mesh as counter and Ag/AgCl (KCl 3M, Bioanalytical systems, the Netherlands) as reference electrode. The precursor solution for the $Mn_xO_y$ deposition contained 0.1 M $MnSO_4$ and 0.5 M $H_2SO_4$. For the Mo-doped $Mn_xO_y$, 100 µM $Na_2MoO_4$ was also added to the electrodeposition bath. The electrodeposition was performed in the potentiostatic mode at 1.7 V/SHE (vs Standard Hydrogen Electrode). To ensure that the equal amount of $Mn_xO_y$ was deposited in each procedure, the loading mass was estimated according to Faraday's law and the applied charge was limited to 13 C. The surface area of the Ti mesh coated with pure or Mo-doped $Mn_xO_y$ catalyst exposed to the electrolyte was 5 x 4 cm (**Figure S1**). Once coated, the anode was thoroughly rinsed with the deionized water and dried under a nitrogen stream.

*2.2. Material characterization*

The surface morphology of the synthesized materials was examined using an ultra-high-resolution field emission scanning electron microscopy (FESEM) (The Magellan 400L, FEI, US). The cross-section images were taken from the cracked layers after bending the samples. X-ray photoelectron spectroscopy (XPS) was performed using PHOIBOS 150 X-ray photoelectron spectrometer (Specs, Germany). The crystal structure of the $Mn_xO_y$ coating could not be determined due to the insufficient intensity of the diffraction peaks, likely as a consequence of the insufficient crystallinity of the coating.

Electroactive surface area of the Ti/TiO$_2$ NTA–Mn$_x$O$_y$ and Ti/TiO$_2$ NTA–Mn$_x$O$_y$-Mo electrodes was estimated by measuring the double layer capacitance observed in the cyclic voltammetry (CV) measurements in 0.1 M NaNO$_3$. Voltametric scans were performed over a potential window of 0.8 – 1 V/SHE and at scan rates of 1 - 5 mV s$^{-1}$. Values for $C_{dl}$ were determined by the linear regression of the current versus the scan rate, according to the following equation:

$$\frac{I_a - I_c}{2} = C_{dl} \nu \qquad (eq.\ 1)$$

where $I_a$ and $I_c$ are the anodic and the cathodic currents observed in the forward and reversed scans (mA), respectively, and $\nu$ is the applied scan rate (mV s$^{-1}$) [22].

Electrical conductivities of the Ti/TiO$_2$ NTA–Mn$_x$O$_y$ and Ti/TiO$_2$ NTA–Mn$_x$O$_y$-Mo electrodes were compared using electrochemical impedance spectroscopy (EIS). The measurements were performed in 0.1 M NaNO$_3$ and applying the potential of 1.4 V/SHE in the frequency range of 0.1 Hz−50 kHz. The results of the measurements were represented in Nyquist plots. Also, electrochemical activities of the synthesized materials were evaluated by performing linear sweep voltammetries (LSVs) in 50 mM NaNO$_3$ electrolyte at 20 mV sec$^{-1}$ scan rate.

### 2.3. Removal of persistent organic contaminants with TiO$_2$ NTA–Mn$_x$O$_y$ and TiO$_2$ NTA–Mn$_x$O$_y$-Mo anodes

Electrochemical experiments were performed in a 100 mL cell divided by a porous glass frit, to avoid the reduction of the contaminants at the counter electrode. The synthesized materials were used as anodes, whereas the Ag/AgCl (3M KCl) and Pt wire (Advent Research Materials, UK) served as the reference and counter electrodes, respectively. All experiments described in this chapter were performed in duplicates, and the results are presented as mean values with their standard deviations.

First, TiO$_2$ NTA–Mn$_x$O$_y$ and TiO$_2$ NTA–Mn$_x$O$_y$-Mo anodes were investigated in the open circuit (OC) experiments to evaluate the intrinsic catalytic activity of the Mn$_x$O$_y$ and determine the impact of Mo doping on this activity. The experiments were performed in 5 mM NaNO$_3$ supporting electrolyte as its conductivity and pH (i.e., 1.3 mS cm$^{-1}$, pH 8.2) are typical for various sources of drinking water (e.g., surface or groundwater). The supporting electrolyte was amended with the mixture of persistent organic contaminants (i.e., ACY, CBZ, DCL, TCL, TMD). The initial concentration of each contaminant was 1 µM to allow their detection by the employed analytical method. Chemical structures and physico-chemical properties of the target contaminants are summarized in **Table S1**.

The impact of the applied anode potential and the effect of Mo doping on the removal of target contaminants were investigated by performing the electrocatalytic experiments using TiO$_2$ NTA-Mn$_x$O$_y$ or TiO$_2$ NTA-Mn$_x$O$_y$-Mo anodes in chronoamperometric mode at 1 V, 1.2 V or 1.4 V/SHE. The potential range tested in this study was established based on the linear sweep voltammetry (LSV), which demonstrated a well-defined peak at 1.2 V/SHE (**Figure S2**). The peak was attributed to the electron transfer between Mn III and Mn IV redox couple [23]. The supporting electrolyte and the concentrations of the target organic contaminants were the same as in the open circuit experiments (i.e., 5 mM NaNO$_3$ and ACY, CBZ, DCL, TCL, TMD at 1 µM each).

The effect of the supporting electrolyte was investigated by performing the experiments in the OC and at different applied anode potentials (i.e., 1 V, 1.2 V or 1.4 V/SHE) using TiO$_2$ NTA–Mn$_x$O$_y$ anode in nitrate, phosphate, and chloride-based supporting electrolytes at the same conductivity (i.e., 1.3 mS cm$^{-1}$), amended with the target contaminants at the same initial concentration of 1 µM each. The pH of the supporting electrolytes was adjusted to 8.2.

Finally, the performance of the TiO$_2$ NTA–Mn$_x$O$_y$-Mo anode was evaluated in real tap water (pH 7.5, σ 0.38 mS cm$^{-1}$, TOC 36.3 mg L$^{-1}$, chloride 33.8 mg L$^{-1}$), amended with the mixture of the target contaminants (initial concentrations of 1 µM). The experiment was performed at 1.4 V/SHE, as this potential yielded the highest electro-oxidation rates of organic contaminants, whereas further increase in potential would compromise the anode stability. To compare the performance of the TiO$_2$ NTA–Mn$_x$O$_y$-Mo anode with the commercially available materials, electrochemical oxidation of target organic contaminants was also performed using Ti/IrO$_2$-Pt anode in the same set-up. Ti/IrO$_2$-Pt was polarized at 1.4 V/SHE in 5 mM NaNO$_3$ (pH 8.2, σ 1.3 mS cm$^{-1}$) amended with the mix of target contaminants.

The electric energy per order (E$_{EO}$, kWh m$^{-3}$) was calculated as follows:

$$E_{EO} = \frac{U \int_0^t I_t dt}{V \log\left(\frac{c_0}{c_t}\right)} \qquad (eq.\ 2)$$

where U is the applied voltage (V), I$_t$ is the current (mA), V represents volume (unit) of the treated electrolyte and c$_0$ and c$_t$ are the initial and average final concentration of the selected contaminants.

Electro-generation of OH• radicals was probed by performing the experiments with terephthalic acid (TA) at the highest tested potential (i.e., 1.4 V/SHE), as this compound is not amenable to direct electrolysis [24]. The electrolyte used in this experiment contained 20 mg L$^{-1}$ TA and 5 mM NaNO$_3$ at pH 8.2. The quasi-steady state OH• concentration was determined as a ratio of pseudo-first rate constant of TA decay ($k_{TA}$, s$^{-1}$) and $k_{TA,HO•}$ [22]. The generation of singlet oxygen was probed by performing the experiments in the electrolyte amended with 5 µM furfuryl alcohol (FA) as well as the selected contaminants. This concentration was equivalent to the total amount of the added target contaminants and selected to ensure the conditions of competitive kinetics, as using scavenging alcohol in excessive amounts in electrochemical process is likely to saturate the anode surface and block the active sites [25]. FA reacts with $^1O_2$ with a known reaction rate constant, $k_{FA,1O2} = 1.2 \times 10^8$ M$^{-1}$ s$^{-1}$.

*2.4 Analytical methods*

The selected organic contaminants were analyzed with a 5500 QTRAP hybrid triple quadrupole-linear ion trap mass spectrometer with a turbo Ion Spray source (Applied Biosystems), coupled to a Waters Acquity Ultra-Performance™ liquid chromatograph (Milford). The details of the target analytical method are given in **Text S1** and **Table S2, S3.** The identification of possible transformation products (TPs) was carried out by Orbitrap™-high-resolution mass spectrometry (HRMS) (Thermo Fisher Scientific Inc), by correlating the accurate *m/z* of all detected compounds and the corresponding fragmentation pattern (**Figure S11-S14**). The analytical method is described in detail in **Text S2**.

The stability of the $Mn_xO_y$ coating was estimated by determining the concentration of the total dissolved manganese. The measurement was performed with inductively coupled plasma-optical emission spectrometry (ICP-OES) (Agilent 5100, Agilent Technologies, US).

Free chlorine and ozone were measured in NaCl (pH 8.2, σ 1.3 mS cm$^{-1}$) and in tap water (pH 7.5, σ 0.38 mS cm$^{-1}$) immediately after sampling, using a diethyl-p-phenylene diamine (DPD) colorimetric method with Chlorine/Ozone/Chlorine dioxide cuvette tests LCK 310 (Hach Lange Spain Sl). Tap water was stirred overnight to ensure a complete absence of chlorine in the beginning of the experiment. The quantification limit for ozone and chlorine was 0.01 mg L$^{-1}$ and 0.05 mg L$^{-1}$, respectively. Hydrogen peroxide ($H_2O_2$) was analyzed by a spectrophotometric method with the quantification limit of 1 μM, using 0.01 M copper (II) sulphate solution and 0.1% w/v 2,9 dimethyl -1,10 – phenanthroline (DMP). In the presence of $H_2O_2$ these compounds form a $Cu(DMP)_2^+$ cation, which yields an adsorption peak at 454 nm [26]. The concentration of TA was measured using HPLC-UV (Agilent Technologies 1200 series) analysis at 239 nm (quantification limit of 1.98 mg L$^{-1}$).

## 3. Results and discussion

### 3.1. Characterization of the TiO$_2$ NTA electrodes coated with Mn$_x$O$_y$, and Mn$_x$O$_y$ doped with Mo

The electrode synthesis procedure yielded a uniform TiO$_2$ nanotubes layer. The nanotubes had an average outer diameter of 80–100 nm, wall thickness of 7-10 nm and length of about 1 μm **(Figure S3a)**. Annealing at 400°C did not impact the structure or morphology of the nanotubes, but it transformed the rutile phase typical for the untreated TiO$_2$ into a pure tetragonal anatase **(Figure S4)** [27]. The Ti 2p spectra obtained with the XPS exhibited two peaks (i.e., at 459.5 eV and 465.2 eV), which can be ascribed to Ti 2p3/2 and Ti 2p1/2 (**Figure 1d**). The spin orbit splitting of 5.8 eV is typical of Ti with the valence state of IV [28].

SEM image of the cross-section of the materials after coating (**Figure S1b**) demonstrates that the electrodeposition does not lead to any visible change in the NTA morphology. Additionally, the crystalline structure of $TiO_2$ NTA remained unchanged, as the diffraction patterns of the $Ti/TiO_2$ NTA samples before and after the coating procedure were similar **(Figure S4)**. As it can be noted from the XPS spectra (**Figure S3c and S3d**), electrodeposition led to a shift of both Ti 2p3/2 and Ti 2p1/2 peaks to lower binding energies (i.e., 458.7 eV and 463.1 eV, respectively). This shift typically indicates an increase in the electron cloud density around Ti, which may occur as a result of the substitution of oxygen in the crystal lattice of $TiO_2$ with a more electronegative atom [29]. Considering that Mn electronegativity (1.55 on Pauling scale) is lower than that of O (3.44 on Pauling scale), this shift suggests a strong interaction between the $Mn_xO_y$ coating and the $Ti/TiO_2$ NTA substrate [30]. The strong interaction between the $Mn_xO_y$ coating and the $TiO_2$ NTA was also confirmed by the SEM images. Anodic electrodeposition enabled the penetration of the MnxOy inside the nanotubes (**Figure S3b**). The filling of the nanotubes was followed by the formation of a uniform $Mn_xO_y$ coating with the thickness of 100 nm on the top of the NTA interlayer. According to the SEM images, pure $Mn_xO_y$ coating was comprised of compact grains (**Figure 1a**). The incorporation of Mo, which was detected by EDX (**Figure 1b**), did not cause any significant change in the $Mn_xO_y$ coating morphology. However, Mo doping slightly increased the relief of the $Mn_xO_y$ coating, which can indicate an increase in the surface area compared to the non-doped sample. The assumption that the incorporation of Mo into the $Mn_xO_y$ lattice leads to a higher electrode surface area was confirmed by measuring the double layer capacitance of the pure and doped $Mn_xO_y$ anodes. Based on the obtained results, Mo doping led to 1.4-fold increase of the electroactive surface area when compared to pure $Mn_xO_y$ (i.e., from 2059 cm$^2$ for Ti/TiO$_2$ NTA–$Mn_xO_y$ to 2883 cm$^2$ for Ti/TiO$_2$ NTA–$Mn_xO_y$-Mo) (**Figure S5**). High electroactive surface areas are likely a consequence of the contributions of pseudo- and chemical capacitance, and capacitance due to residual charge-transfer processes to the measurement [31].

Further increase of the $MoO_4^{2-}$:$Mn^{2+}$ ratio in the electrodeposition bath (i.e., from 1:1000 to 1:100 $MoO_4^{2-}$ : $Mn^{2+}$) produced an uneven coating with poor coverage of the TiO$_2$ NTA interlayer (**Figure S6**). This can be explained by the enhanced incorporation of Mo into $Mn_xO_y$, which leads to structural instability and even collapse of the host metal lattice [32]. Therefore, excessive $MoO_4^{2-}$ concentration in the precursor solution is unfavorable for the synthesis of a stable and homogeneous Mo-doped $Mn_xO_y$ coating.

The chemical states of the elements present in the synthesized coatings were investigated with the XPS analysis (**Figure 2**). The valence state of the $Mn_xO_y$ catalyst was determined based on the value of the Mn 3s spin orbit splitting, given that this approach was reported to be more reliable compared to the deconvolution of the Mn 2p core level spectra. However, the analysis of the Mn 2p binding energy region can still provide valuable information on the chemical state of the $Mn_xO_y$ catalyst. The Mn 2p spectra demonstrated two well defined peaks at 653.9 eV and ~ 642.2 eV, with the spin orbit splitting of ~11.7 eV in both pure and doped $Mn_xO_y$ samples (**Figure 2a**). The asymmetry of the Mn 2p$_{3/2}$ peak is an indicator of the presence of Mn in the mixed valence states of Mn III and Mn IV. However, considering the splitting of 4.9 eV observed in the Mn 3s core level, the major oxidation state of the Mn sites is IV [33]. Indeed, this oxidation state is common for the $Mn_xO_y$ coatings synthesized via the anodic electrodeposition in acidic solutions, and is formed according to the following reaction [34]:

$$\text{Mn}^{2+}_{(aq)} + 2\text{H}_2\text{O} \rightarrow \text{MnO}_{2(s)} + 4\text{H}^+ + 2\text{e}^- \qquad (\text{eq. 3})$$

The splitting of the Mn 3s spin orbit was increased (i.e., from 4.9 eV to 5.1 eV) once the $\text{Mn}_x\text{O}_y$ was doped with Mo (i.e., from 4.9 eV in the pure $\text{Mn}_x\text{O}_y$ to 5.1 eV in the Mo-doped sample) (**Figure 2d**). The ΔE value observed in the Mo-doped samples corresponds to an average oxidation state of 3.3, according to a linear relationship between the Mn oxidation state and the ΔE value reported in literature (**Figure S7**) [35]. Thus, the incorporation of Mo into the $\text{Mn}_x\text{O}_y$ lattice increases the presence of Mn sites with the valence state of III, leading to a lower overall oxidation state of the Mn catalyst.

The XPS confirmed further the results of the EDX analysis, clearly indicating the presence of Mo in the doped samples (**Figure 2f**). The obtained Mo 3d spectra exhibited two peaks ascribed to Mo $3d_{5/2}$ and $3d_{3/2}$ components, with the spin orbit splitting of ~3.2 eV. The asymmetry of these core level spectra suggests that Mo is present in the $\text{Mn}_x\text{O}_y$ metal lattice in several valence states. Hence, both Mo $3d_{5/2}$ and $3d_{3/2}$ were deconvoluted into two peaks. The peaks observed at 231.8 eV and 235 eV were ascribed to Mo with the valence state of IV, while the other pair at 232.2 eV and 235.5 eV was assigned to Mo VI [36]. The atomic percentage of Mo in the $\text{Mn}_x\text{O}_y$ lattice was estimated at 12.5%. Furthermore, the O1s XPS spectrum for both coated and non-coated materials exhibits a singlet peak at 529.6 – 530 eV, attributed to lattice oxygen with a smaller shoulder at 530.9 – 531.8 eV that represents non-lattice oxygen or oxygen vacancy [28]. The third peak was assigned to the residual structural water (**Figure 2c and 2g, Table S4**). The atomic percentage of the non-lattice oxygen increased from 36.35% for pure $\text{Mn}_x\text{O}_y$ to 42.7% for Mo-doped $\text{Mn}_x\text{O}_y$, indicating the increase of oxygen vacancies in the $\text{Mn}_x\text{O}_y$ lattice induced by the presence of Mo.

Based on the results described above, a possible mechanism for the formation of the Mo-doped $\text{Mn}_x\text{O}_y$ coating can be proposed. The process starts with the electrooxidation of aqueous $\text{Mn}^{2+}$, which leads to the precipitation of $Mn_xO_y$ at the anode surface and accumulation of protons (eq. 4). The acidic environment in the vicinity of the anode promotes the conversion of $MoO_4^{2-}$ into polymolybdate, which becomes dehydrated and forms co-precipitate with the $Mn_xO_y$, according to the following reaction [35]:

$$7MoO_4^{2-}{}_{(aq)} + 8H^+ \rightarrow Mo_xO_{y\ (s)} + H_2O \quad\quad\quad (eq.\ 4)$$

Once the co-precipitate is formed, $Mo^{6+}$ is partially reduced to $Mo^{4+}$ through Mn centers. This reduction can be induced by the weakly bonded lattice oxygen through charge compensation mechanism and random potentials around the metal lattice. $Mo^{6+}$ reduction can also be assisted by the oxygen vacancies, according to the following reaction [36]:

$$Mo^{6+}{}_{(s)} + O^{2-} \rightarrow Mo^{4+}{}_{(s)} + 0.5\ O_2 \quad\quad\quad (eq.\ 5)$$

The XRD diffraction pattern of the coated sample did not contain any signal of $Mn_xO_y$ crystalline phases (**Figure S2**). The absence of the $Mn_xO_y$ signal in the XRD patterns can be explained by the insufficient crystallinity of the coating, as the electrodeposited $Mn_xO_y$ is typically characterized by an amorphous structure [53]. In summary, our results demonstrate that the incorporation of Mo into the electrodeposited $Mn_xO_y$ induces significant structural modifications of the coating. The effect of these modification on the catalytic activity and electrocatalytic performance of the obtained anodes is discussed below.

### 3.2. Electrocatalytic performance of the Ti/TiO$_2$ NTA–Mn$_x$O$_y$ and Mn$_x$O$_y$-Mo electrodes

The catalytic activity of the Ti/TiO$_2$ NTA–Mn$_x$O$_y$ anode materials was first evaluated in the OC experiments (**Figure 3**). Even without the application of any potential, the non-doped Ti/TiO$_2$ NTA–Mn$_x$O$_y$ was capable of low to moderate contaminant removal (i.e., 12.3% of ACY, 6% of CBZ, 9.2% DCL, 41.6% of TCL, 7.1% TMD). Indeed, the elimination of certain organic contaminants (e.g., ACY, CBZ, DCL, TCL, TMD) on manganese oxides achieved though oxidative transformation or adsorption has been previously reported [37-39]. Manganese oxide catalyst is a hydrated oxide, with the surface covered with hydroxyl groups, formed by the dissociative chemisorption of water molecules. Depending on the pH of the solution, hydroxyl group can become protonated by releasing an OH$^-$ ion and generate a positively charged site (i.e., $\equiv$MnOH$_2^+$) or yield a proton and a negatively charged site (i.e., $\equiv$MnO$^-$) [40]:

$$\equiv\text{MnOH} \rightleftarrows \equiv\text{MnOH}_2^+ + \text{OH}^- \qquad (eq.\ 6)$$

$$\equiv\text{MnOH} \rightleftarrows \equiv\text{MnO}^- + \text{H}^+ \qquad (eq.\ 7)$$

Considering that the point of zero charge (PZC) of manganese oxide is 4, at circumneutral pH typical for drinking water the surface of manganese oxide is covered by undissociated hydroxyl groups (i.e., $\equiv$MnOH) and negatively charged sites (i.e., $\equiv$MnO$^-$) [40].

The simplest model that describes the oxidation of organic contaminants on metal oxides is a surface site-binding model, which involves the formation of a precursor complex between the metal center and the organic molecule [41]. Positively charged (TMD) can easily form a complex with the negatively charged $\equiv$MnO$^-$ sites. Adsorption of the negatively charged organic molecules (i.e., DCL) is hindered by the electrostatic repulsion, however, they still can bind with the undissociated acidic cites (i.e., $\equiv$MnOH) [41, 42]. Complex formation between the neutral molecules (i.e., ACY, CBZ, TCL) and the catalyst can occur at both undissociated and negatively charged sites. Next, organic pollutant is oxidized whereas Mn IV is reduced to Mn II. The resulting Mn II is then released into the electrolyte in the form of aqueous Mn$^{2+}$ [39, 41]. In addition to oxidative transformation, positively charged molecules can simply be retained by the negatively charged Mn$_x$O$_y$ catalyst, as a result of the electrostatic attraction [41]. To determine whether the contaminants were transformed or simply adsorbed onto the $Mn_xO_y$ coating, the catalyst was completely dissolved with ascorbic acid after the experiment. None of the initial contaminants could be detected upon the coating dissolution, meaning that their removal was achieved only by oxidation. This was further supported by the release of the dissolved $Mn^{2+}$ detected at the end of the OC experiment (i.e., 5.86±0.05 mg $L^{-1}$), which indicated the reduction of the $Mn_xO_y$ catalyst due to its participation in the contaminant oxidation (**Table S5**).

Application of low anodic potential (i.e., 1 V/SHE) significantly improved the removal efficiencies and the oxidation kinetics of ACY, CBZ, DCL and TCL at the $Ti/TiO_2$ NTA–$Mn_xO_y$. Anodic polarization reduced the repulsive forces between negatively charged molecules (e.g., DCL) and manganese oxide surface observed in OC experiment, facilitating the formation of the precursor complexes. Furthermore, the application of potential can accelerate the electron transfer within the complex, thus leading to higher oxidation rates of all model organic contaminants. Increase in the applied anode potential above 1 V/SHE had positive impact on the electro-oxidation of ACY, CBZ, DCL and TCL, and enabled their almost complete removal within 2 hours (**Table S6**). The removal efficiency of TMD at the highest tested potential (i.e., 1.4/SHE) was estimated at 63.9%. Limited removal efficiency of TMD could be attributed to its stable molecular structure. Moreover, positive potential may impede diffusion of positively charged TMD molecule towards the anode surface.

In addition to enhancing the oxidation of pollutants, anodic polarization further stabilizes the non-doped $Mn_xO_y$ coating. Applied potential prevents the release of $Mn^{2+}$ into the solution by regenerating, i.e., re-oxidizing it to its initial oxidation state [43]. The application of potential to the $Ti/TiO_2$ NTA –$Mn_xO_y$ anode reduces the catalyst loss from 16.2% in the OC, to 11.5% at 1 V/SHE (**Table S7**). Increase in the applied anode potential to 1.4 V/SHE completely prevents the dissolution of the $Mn_xO_y$. However, further increase of potential (i.e., above 1.4V/SHE) leads to $Mn_xO_y$ catalyst corrosion due to its oxidation to permanganate [10]. Therefore, continuous regeneration of the $Mn_xO_y$ catalyst achieved through anodic polarization of the Ti/TiO$_2$ NTA– $Mn_xO_y$ electrode accelerates the oxidation of the organic contaminants and ensures a complete stability of the electrode.

Doping of the $Mn_xO_y$ with Mo substantially improved the removal efficiency of the contaminants already in the OC experiments, in particular ACY (i.e., from 12.3% to 51.8%), DCL (i.e., from 9.2% to 64.7%) and TCL (i.e., from 41.6% to 83.5% for pure and Mo-doped $Mn_xO_y$, respectively) (**Figure 3**). In case of DCL, improved removal can arise from reduced electrostatic repulsion between the negatively charged molecule and the catalyst surface. As was mentioned previously, incorporation of Mo induces the occurrence of the oxygen vacancies in the host lattice of the $MnO_2$. When an oxygen vacancy is formed, the neighboring Mn atoms become positively charged, thus providing more sites available for the interaction with the negatively charged DCL molecule [44].

The presence of oxygen vacancies was reported to enhance the catalytic activity of metal oxides [47]. Abundant oxygen vacancies cause weakening of the Mn–O bonds, which enhances the mobility of oxygen within the metal lattice [45]. Oxidation of the organic molecules on metal oxides can be described through Mars-Van Krevelen mechanism, a more complex model compared to site-binding one as it also considers the role of lattice oxygen in the oxidation process. According to Mars-Van Krevelen model, high oxygen mobility within the metal oxide facilitates its diffusion through the catalyst and the subsequent reaction with the organic molecule, thus leading to a higher reaction rate. Besides, the presence of oxygen vacancies enables the formation of active oxygen species at the metal surface. Molecular oxygen can chemisorb into the defective sites in the metal lattice even under ambient conditions, and remain bound to the surface in superoxo- ($O_2^-$) or peroxo-like ($O_2^{2-}$) state, further accelerating the catalytic oxidation of organic contaminants [46, 47]. Lack of structural defects like in case of the anode coated with pure $Mn_xO_y$ limits the generation of these species, as in this case $O_2$ cannot form a bond with the metal oxide and remains only physisorbed at its surface [48].

Higher catalytic activity of the Ti/TiO$_2$ NTA–Mn$_x$O$_y$-Mo material could be also attributed to the mixed valence state of the catalyst. Mo doping reduces the overall oxidation state of the Mn$_x$O$_y$ by increasing the number of Mn III sites. In addition to this, Mn$_x$O$_y$ doping with Mo incorporates Mo IV and Mo VI sites within the host lattice. This mixed valence state leads to the formation of redox couples between the Mn and Mo species, thus increasing the oxidizing power of the catalyst. Electron transfer between the Mn and Mo redox couples was also confirmed by the increase of the OC voltage, i.e., from 0.72 V/SHE at Ti/TiO$_2$ NTA–Mn$_x$O$_y$ to 0.8 V/SHE at Ti/TiO$_2$ NTA–Mn$_x$O$_y$-Mo. Furthermore, unlike the OC experiments with pure Mn$_x$O$_y$, $Mn^{2+}$ release from the Mo-doped Mn$_x$O$_y$ was negligible (**Table S5**), meaning that the mixed valence state effectively regenerated the reduced Mn II back to its initial state and prevented its release as $Mn^{2+}$ in the presence of organics [49]. In our previous study [43], anodic polarization at low potentials was needed to prevent the release of $Mn^{2+}$ in the presence of reactants. Here, doping with Mo greatly prevented the dissolution of the Mn$_x$O$_y$ coating, thus ensuring the material stability even in the OC.

In electrochemical experiments at 1 - 1.4V/SHE of applied anode potential, Mo-doped Mn$_x$O$_y$ electrode clearly demonstrated higher removal efficiencies and rates of organic pollutants compared to the pure Mn$_x$O$_y$ coating (**Figure 3, Table S6, Table S8**). This improved performance of the Ti/TiO$_2$ NTA–Mn$_x$O$_y$-Mo anode could arise from its higher electrochemically active surface area. However, the removal rates normalized to the surface area were still higher at Mo-doped material, meaning that the higher electroactive surface was not the only reason for better performance (**Table S8**) [36]. Superior electrocatalytic activity of the Mo-doped material can also be attributed to its higher electrical conductivity. **Figure S8a** depicts the Nyquist diagrams for Ti/TiO$_2$ NTA–Mn$_x$O$_y$ or Mn$_x$O$_y$-Mo anodes obtained in 100 mM NaNO$_3$ electrolyte at 1.4 V/SHE. Based on this diagram, charge transfer resistance ($R_{ct}$) for pure Mn$_x$O$_y$ can be estimated at 3.1 Ω, while for the Mo-doped material it was reduced to 1.1 Ω. Lower resistance of the Mn$_x$O$_y$-Mo coating means that the incorporation of Mo facilitates the electron transfer within the precursor complex between the catalyst and the organic molecule, which can enhance the kinetics of the oxidation reaction. The enhanced electrochemical performance of the Mn$_x$O$_y$ anode doped with Mo was further supported by the LSV (**Figure S8b**), which displayed higher current density compared with the pure Mn$_x$O$_y$. Excellent conductivity of the Mo-doped Mn$_x$O$_y$ electrode can be attributed to the synergistic effect of the interaction between Mo and Mn redox couples and higher content of oxygen vacancies, which accelerate the electron transfer.

Improved electrocatalytic activity of the Ti/TiO$_2$ NTA–Mn$_x$O$_y$-Mo anode may also be caused by the chemisorbed oxygen dissociation, which cannot occur at the anode coated with undoped Mn$_x$O$_y$ due to the lack of oxygen vacancies in the metal lattice. Once the Ti/TiO$_2$ NTA–Mn$_x$O$_y$-Mo anode is polarized and sufficient charge transfer is achieved, chemisorbed oxygen can dissociate into two surface bound oxygen atoms, which in some cases are more active towards the oxidation of certain molecules compared to the dioxygen species (i.e., O$_2^-$ or O$_2^{2-}$) [48].

Enhanced removal efficiencies of the target contaminants achieved with the application of anodic potential can result from oxygen activation at the Mn$_x$O$_y$-Mo catalyst under the influence of the electric field [25]. To clarify the contribution of the major oxidant species (OH$^•$, H$_2$O$_2$, O$_3$) in the overall removal mechanism, their formation at the Ti/TiO$_2$ NTA–Mn$_x$O$_y$-Mo anode was monitored in the absence of contaminants at 1.4 V/SHE, however, none of the oxidants could be detected. Singlet oxygen can also be generated at the $Mn_xO_y$-Mo anode [50]. To determine whether the generation of $^1O_2$ occurred, experiments were conducted in the presence of 5 μM FA to enable competitive kinetics and avoid blockage of the $Mn_xO_y$ active sites with alcohol [25]. The effect of FA presence on the removal of contaminants at the Ti/TiO$_2$ NTA–$Mn_xO_y$-Mo anode at 1.4 V/SHE was negligible, excluding the participation of $^1O_2$ in the oxidation (**Figure S9**). Thus, the increase of removal efficiencies observed at higher potential occurred as a result of an enhanced electron transfer and re-oxidation of the $Mn_xO_y$-Mo catalyst and did not involve the generation of oxidant species (i.e., OH$^•$, $H_2O_2$, $O_3$ or $^1O_2$).

Superior electrochemical performance of the Ti/TiO$_2$ NTA–$Mn_xO_y$-Mo anode led to an enhanced oxidation of persistent organic contaminants, and lower energy requirements of the system. The $E_{EO}$ required for the removal of contaminants is lower in a system equipped with Mo-doped $Mn_xO_y$ anode (**Figure S10**). For example, $E_{EO}$ required for CBZ removal using the $Mn_xO_y$ and $Mn_xO_y$-Mo anode was 8.5 and 7.2 W h m$^{-3}$, respectively. However, the coating is not the only anode component that helps to lower the energy requirements. The Ti/TiO$_2$ NTA substrate also plays an important role in the excellent electrochemical performance of the system. The TiO$_2$ NTA interlayer promotes the charge transfer between the $Mn_xO_y$ coating and the Ti support, thus improving the electrode stability and activity [43, 51]. As a result of the synergy between the electrocatalytically active coating and the conductive and stable substrate, Ti/TiO$_2$NTA–$Mn_xO_y$-Mo anode offers significantly improved removal efficiency and oxidation kinetics (**Figure 4a and 4b**). More importantly, due to the catalytic properties of the $Mn_xO_y$-Mo coating, oxidation of contaminants can be achieved at low applied potentials at which commercial MMO anodes typically have limited electrocatalytic performance. In the experiments run with the commercial

Ti/IrO$_x$-Pt anode at 1.4 V/SHE, limited removal efficiencies were observed (i.e., 24% for ACY, 15% for CBZ, 0% for DCF, 45% for TCL and 12% for TMD). In addition to poor removal efficiency, the energy required for the removal of target organic contaminants at Ti/IrO$_x$-Pt anode is several orders of magnitude higher compared to both Ti/TiO$_2$ NTA–Mn$_x$O$_y$ and Ti/TiO$_2$ NTA–Mn$_x$O$_y$-Mo developed in this study **(Figure 4c)**. For instance, E$_{EO}$ for DCL removal on Ti/IrO$_x$-Pt in the same set up was calculated at 3.4 kW h m$^{-3}$, while an electrochemical system equipped with Ti/TiO$_2$ NTA–Mn$_x$O$_y$ or Ti/TiO$_2$ NTA–Mn$_x$O$_y$-Mo anodes only needed 6.8 x 10$^{-3}$ and 2.6 x 10$^{-3}$, respectively.

### *3.3. The impact of the supporting electrolyte*

The impact of anions, typically present in water, was determined by performing the experiments in NaNO$_3$, Na$_3$PO$_4$ and NaCl supporting electrolytes with similar conductivity and pH (i.e., 1.3 mS cm$^{-1}$ and pH 8.2). Oxidation of contaminants was hampered in phosphate-based electrolyte compared to NO$_3^-$ or Cl$^-$, in both OC and chronoamperometric experiments (**Figure 5** and **Table S9**). The negative effect of phosphate on the oxidation of contaminants can be explained by its tendency to adsorb onto manganese oxides [52]. The surface hydroxyl groups on the manganese oxide react with the phosphate ions to form the outer sphere complexes, which can block or hinder the formation of the precursor complexes between the organic molecules and the Mn$_x$O$_y$ active sites:

$$\equiv MnOH + 2H^+ + HPO_4^{2-} \rightleftarrows \equiv MnOH_2^+ - HPO_4^{2-} \quad \text{(eq. 8)}$$

Also, negative charge generated by the presence of phosphate surface complexes can enhance the electrostatic repulsion between the negatively charged organic molecules and the anode surface. Moreover, anodic polarization of the Mn$_x$O$_y$ can promote the uptake of the phosphate anions through electrosorption, thus further aggravating the catalyst inactivation [53].

The presence of Cl⁻ had no effect on the removal rates of the contaminants compared to the nitrate-based electrolyte (**Figure 5** and **Table S8**). Given the high reported reactivity of DCL and TMD with free chlorine, this implies that the Ti/TiO$_2$ NTA–Mn$_x$O$_y$-Mo anode suppressed chloride oxidation even at the highest tested potential (i.e., 1.4 V/SHE) [54, 55]. Indeed, the formation of free chlorine in the absence of the contaminants at this potential was negligible (**Table S10**). Although chlorine evolution could accelerate the removal of the contaminants through their indirect oxidation, it would also lead to the formation of their potentially toxic chlorinated by-products.

### *3.4. Experiments with tap water*

The performance of the Ti/TiO$_2$ NTA–Mn$_x$O$_y$-Mo anode was investigated in real tap water amended with the target contaminants (**Figure 6**). The removal efficiencies of ACY, CBZ and DCL were only slightly decreased in tap water compared with the experiments with the supporting electrolyte of higher conductivity. However, in the case of TCL and TMD, the decrease of the reaction rate was more prominent. As was mentioned previously, TMD diffusion towards the anode surface is impeded by the electrostatic repulsion between the positively charged molecule and the anode surface. Considering that the low conductivity of the tap water (i.e., 0.38 mS cm$^{-1}$) thickens the electrical double layer, the diffusion of the TMD molecule becomes more limited compared to other target pollutants that were uncharged (i.e., ACY, CBZ) or negatively charged (i.e., DCL), which leads to significantly lower removal rates [56, 57]. In the case of TCL, given that its pKa is 7.9, slightly lower pH of tap water (i.e., pH 7.5) compared to NaNO$_3$ supporting electrolyte (i.e., pH 8.2) favored the protonation of TCL. Phenolic form of TCL is characterized by the lower reactivity compared to its deprotonated form, which could explain the decrease of the removal rate. Oxidation of the selected organic contaminants can also be hampered by the passivating effect of phosphate, or the competition imposed by the presence of another ions (e.g., organic matter) [41, 52].

The electro-generation of free chlorine was investigated in the employed tap water (33.8 mg/L of $Cl^-$) at the Ti/TiO$_2$ NTA–Mn$_x$O$_y$-Mo and Ti/IrO$_x$-Pt anodes at 1.4 V/SHE, without the addition of contaminants. The concentration of the residual free chlorine measured in system equipped with Ti/TiO$_2$ NTA–Mn$_x$O$_y$-Mo anode was close to the limit of detection (i.e., 0.05 mg L$^{-1}$), while polarization of Ti/IrO$_x$-Pt resulted in substantial increase of free chlorine (i.e., 0.132 mg L$^{-1}$) at the end of the experiment. Therefore, the Ti/TiO$_2$NTA–Mn$_x$O$_y$-Mo anode displayed a much lower activity towards the chlorine evolution reaction. This performance, coupled with its oxidizing power at low applied potential, represents an important advantage over the conventional MMO anode (i.e., Ti/IrO$_x$-Pt).

### 3.5. Identification of the transformation products and degradation pathways of carbamazepine

High resolution mass spectrometry technique identified four transformation products of CBZ degradation at the Ti/TiO$_2$NTA–Mn$_x$O$_y$-Mo anode at 1.4 V/SHE. Possible TPs are listed in the **Table S11** [58-63]. The identified TPs imply that the degradation of CBZ occurred through the formation of a precursor complex between CBZ and the active site at the anode surface, followed by the electron transfer. Electron transfer enabled the oxidation of the olefinic double bond (C10-C11) on the central heterocyclic ring, characterized by the high reactivity [64]. As a result of the oxidation, olefinic carbon atom yields alcohols, aldehydes, carboxylic acid, and other groups [61, 65]. One possible pathway for the oxidation of CBZ involves intramolecular cyclization [61, 63, 65]. In this case, the formation of the precursor complex and electron transfer leads to electrophilic aromatic substitution in the CBZ molecule. As a result, aromatic ring becomes contracted, yielding

**CBZ-266a** [62, 63]**.** This TP can further transform into **CBZ-223** due to the hydrolysis of the urea group, induced by the presence of the electron-withdrawing carbonyl moieties in the central ring. Alternatively, the complex between CBZ and Mn could overcome hydrolysis leading to the cis-diol formation, followed by the oxidation to yield a keto group **(CBZ-266 b)** [61-63]. Oxidation of CBZ can also occur as result of its interaction with the lattice or surface oxygen groups of the $Mn_xO_y$-Mo catalyst [45]. Addition of oxygen to the olefinic double bond produces a ketone intermediate. This intermediate promotes the hydrolysis of the urea group, further rearrangement reaction and the eventual cleavage of an aldehyde group, thus leading to the formation of **CBZ 179** [66].

According to the identified TPs, the reactivity of the unsaturated carbon bonds in the CBZ molecule was relatively high, which made them susceptible to a reaction with the $Mn_xO_y$-Mo catalyst. The evolution profiles of the major TPs were depicted as a ratio of their peak areas to the area of the spiked CBZ as a function of the treatment time (**Figure 8**). Generation of TPs occurred rapidly starting from the first minutes of the treatment (**Figure 8**). Although **CBZ-179** and **CBZ-223** demonstrated relatively low intensity in comparison with CBZ, they remained persistent during the entire experiment. This trend implies that CBZ oxidation via pathways generating these TPs was not favored. The amount of **CBZ-266(b)** started to decrease after 60 min, reaching negligible values by the end of the experiment. The absence of **CBZ-266(b)** after 90 min (**Figure 8**) can be attributed to the subsequent transformation of **CBZ-266(b)** by the hydroxylation or hydration reactions [58]. The concentration of the major TP **CBZ-266 (a)** was gradually increased during the experiments, implying the pronounced stability of this product to further oxidation. The persistence of this TP is a result of its highly stable structure achieved through high degree of delocalization [58].

## 4. Conclusions

The Ti/TiO$_2$NTA–Mn$_x$O$_y$ anode demonstrated moderate removal (6-41.6%) of the organic pollutants through oxidative transformation even in the absence of anodic polarization. Once the Mn$_x$O$_y$ catalyst was doped with Mo, the removal efficiency and oxidation kinetics of the contaminants were drastically enhanced. The improved performance of the Mo-doped anode can be attributed to two factors. First, the incorporation of Mo into the Mn$_x$O$_y$ increases the number of oxygen vacancies in the lattice. Oxygen vacancies are surrounded by a positive charge density, which reduces the repulsive forces between the negatively charged organic molecules (i.e., DCL) and the negatively charged Mn$_x$O$_y$ catalyst, thus facilitating the formation of the precursor complex. Also, presence of the oxygen vacancies accelerates oxidation of positively charged (i.e., TMD) or non-charged molecules (i.e., ACY, CBZ, TCL) by enhancing the mobility of the lattice oxygen and enabling the formation of the surface bound active oxygen species (i.e., $O_2^-$ or $O_2^{2-}$). Second, the incorporation of Mo allowed the formation of redox couples between the Mn and Mo species (Mn IV/Mn III and Mo IV/IV), which increased the oxidative power of the catalyst. Furthermore, the mixed valence state of the catalyst achieved by the incorporation of Mo effectively stabilized the Mn$_x$O$_y$ coating by preventing its reductive dissolution, which is an important advantage over undoped manganese oxide-based coatings for practical applications of this electrode material, as it does not require constant application of current.

The Ti/TiO$_2$ NTA–Mn$_x$O$_y$ and Ti/TiO$_2$ NTA–Mn$_x$O$_y$-Mo anode polarization at 1 V/SHE enhanced the oxidation of persistent organic contaminants compared to the OC experiments by promoting the electron transfer within the precursor complex, and enabling a continuous reoxidation of the reduced $Mn_xO_y$ catalyst. In the case of the non-doped $Ti/TiO_2$ NTA–$Mn_xO_y$ anode, it helped to stabilize the $Mn_xO_y$ catalyst, preventing the release of $Mn^{2+}$. Removal efficiencies and oxidation kinetics of the contaminants on the Mo-doped anode exceeded those observed at the undoped $Mn_xO_y$ because of the superior electrochemical characteristics of the $Mn_xO_y$-Mo anode, e.g., lower charge transfer resistance and higher electroactive surface area. Furthermore, oxygen chemisorbed into oxygen vacancies can dissociate into two surface bound oxygen atoms under the anodic polarization, which can further accelerate electrooxidation or the organic contaminants. Increase in the anode potential from 1 V/SHE to 1.2 - 1.4 V/SHE further enhanced the removal efficiencies, reaching almost complete (i.e., >90%) removal of the selected contaminants at both anodes. Effective removal of contaminants was also achieved in the low conductivity tap water. However, the removal efficiencies of the positively charged TMD was lowered, likely due to the more pronounced electrostatic repulsion with the anode surface due to the increased thickness of the double layer. Complete removal of positively charged contaminants would require longer treatment times; nevertheless, considering the low energy invested in the performed experiments (1.2 - 15.4 W h m$^{-3}$), anodic oxidation using $Ti/TiO_2$ NTA–$Mn_xO_y$-Mo anode may still be economically feasible for the removal of these pollutants.

The $Ti/TiO_2$ NTA–$Mn_xO_y$-Mo anode developed in this study enabled rapid and effective removal of the persistent organic contaminants and outperformed the commercial MMO electrode (i.e., $Ti/IrO_x$-Pt) in terms of the contaminant removal efficiency, while reducing the energy input requirements for three orders of magnitude, e.g., from 1.5 kW h m$^{-3}$ on $Ti/IrO_x$-Pt to 5.1 W h m$^{-3}$ $Ti/TiO_2$ NTA–$Mn_xO_y$-Mo). Also, the $Ti/TiO_2$ NTA–$Mn_xO_y$-Mo anode effectively minimized the electro-generation of chlorine, thus overcoming this major limitation of the conventional anode materials. Hence, excellent activity of the $Ti/TiO_2$ NTA–$Mn_xO_y$-Mo at potential as low as 1.4

V/SHE, its complete stability even in the absence of anode polarization and its ability to minimize the chlorine evolution suggests that electrocatalytic system based on low-cost Ti/TiO$_2$ NTA–Mn$_x$O$_y$-Mo anodes may be an attractive approach for the removal of persistent organic contaminants from low conductivity solutions.

## Acknowledgments

The authors would like to acknowledge ERC Starting Grant project ELECTRON4WATER (Three-dimensional nanoelectrochemical systems based on low-cost reduced graphene oxide: the next generation of water treatment systems), project number 714177. ICRA researchers thank funding from CERCA program.

## References

- [1] R. Roson, R. Damania, The macroeconomic impact of future water scarcity: An assessment of alternative scenarios, Journal of Policy Modeling, 39 (2017) 1141-1162 https://doi.org/10.1016/j.jpolmod.2017.10.003.
- [2] K. Rabaey, T. Vandekerckhove, A.V. de Walle, D.L. Sedlak, The third route: Using extreme decentralization to create resilient urban water systems, Water Research, 185 (2020) 116-276 https://doi.org/10.1016/j.watres.2020.116276.
- [3] C.A. Martínez-Huitle, M.A. Rodrigo, I. Sirés, O. Scialdone, A critical review on latest innovations and future challenges of electrochemical technology for the abatement of organics in water, Applied Catalysis B: Environmental, 328 (2023) 122-430 https://doi.org/10.1016/j.apcatb.2023.122430.
- [4] A. Kumar, S.-Y. Pan, Opportunities and challenges of electrochemical water treatment integrated with renewable energy at the water-energy nexus, Water-Energy Nexus, 3 (2020) 110-116 https://doi.org/10.1016/j.wen.2020.03.006.
- [5] J.T. Jasper, Y. Yang, M.R. Hoffmann, Toxic Byproduct Formation during Electrochemical Treatment of Latrine Wastewater, Environmental Science & Technology, 51 (2017) 7111-7119 https://doi.org/10.1021/acs.est.7b01002.
- [6] B.P. Chaplin, Chapter 17 - Advantages, Disadvantages, and Future Challenges of the Use of Electrochemical Technologies for Water and Wastewater Treatment, Electrochemical Water and Wastewater Treatment, (2018) 451-494 https://doi.org/10.1016/B978-0-12-813160-2.00017-1.

- [7] J. Radjenovic, D.L. Sedlak, Challenges and Opportunities for Electrochemical Processes as Next-Generation Technologies for the Treatment of Contaminated Water, Environmental Science & Technology, 49 (2015) 11292-11302 https://doi.org/10.1021/acs.est.5b02414.
- [8] C. Minke, M. Suermann, B. Bensmann, R. Hanke-Rauschenbach, Is iridium demand a potential bottleneck in the realization of large-scale PEM water electrolysis?, International Journal of Hydrogen Energy, 46 (2021) 23581-23590 https://doi.org/10.1016/j.ijhydene.2021.04.174.
- [9] M. Shestakova, M. Sillanpää, Electrode materials used for electrochemical oxidation of organic compounds in wastewater, Reviews in Environmental Science and Bio/Technology, 16 (2017) 223-238 https://doi.org/10.1007/s11157-017-9426-1.
- [10] J. Melder, P. Bogdanoff, I. Zaharieva, S. Fiechter, H. Dau, P. Kurz, Water-Oxidation Electrocatalysis by Manganese Oxides: Syntheses, Electrode Preparations, Electrolytes and Two Fundamental Questions, Zeitschrift für Physikalische Chemie, 234 (2020) 925-978 https://doi.org/10.1515/zpch-2019-1491.
- [11] B. Shao, H. Dong, G. Zhou, J. Ma, V.K. Sharma, X. Guan, Degradation of Organic Contaminants by Reactive Iron/Manganese Species: Progress and Challenges, Water Research, 221 (2022) 118765 https://doi.org/10.1016/j.watres.2022.118765.
- [12] J.G. Vos, T.A. Wezendonk, A.W. Jeremiasse, M.T.M. Koper, MnOx/IrOx as Selective Oxygen Evolution Electrocatalyst in Acidic Chloride Solution, Journal of the American Chemical Society, 140 (2018) 10270-10281 https://doi.org/10.1021/jacs.8b05382.
- [13] Z. Hu, M. Chen, H. Zhang, L. Huang, K. Liu, Y. Ling, H. Zhou, Z. Jiang, G. Feng, J. Zhou, Stabilization of layered manganese oxide by substitutional cation doping, Journal of Materials Chemistry A, 7 (2019) 7118-7127 https://doi.org/10.1039/C8TA12515E.
- [14] F. Sabaté, M.J. Sabater, Recent Manganese Oxide Octahedral Molecular Sieves (OMS–2) with Isomorphically Substituted Cationic Dopants and Their Catalytic Applications, Catalysts, 11 (2021) 1147 https://doi.org/10.3390/catal11101147.
- [15] W. Wei, X. Cui, W. Chen, D.G. Ivey, Manganese Oxide-Based Materials as Electrochemical Supercapacitor Electrodes, ChemInform, 42 (2011) 1697-1721 https://doi.org/10.1039/C0CS00127A.
- [16] N. Sergienko, J. Radjenovic, Manganese oxide-based porous electrodes for rapid and selective (electro)catalytic removal and recovery of sulfide from wastewater, Applied Catalysis B: Environmental, 267 (2020) 118-608 https://doi.org/10.1016/j.apcatb.2020.118608.
- [17] A. Massa, S. Hernández, S. Ansaloni, M. Castellino, N. Russo, D. Fino, Enhanced electrochemical oxidation of phenol over manganese oxides under mild wet air oxidation conditions, Electrochimica Acta, 273 (2018) 53-62 https://doi.org/10.1016/j.electacta.2018.03.178.
- [18] Y. Yang, P. Zhang, K. Hu, P. Zhou, Y. Wang, A.H. Asif, X. Duan, H. Sun, S. Wang, Crystallinity and valence states of manganese oxides in Fenton-like polymerization of phenolic pollutants for carbon recycling against degradation, Applied Catalysis B: Environmental, 315 (2022) 121-593 https://doi.org/10.1016/j.apcatb.2022.121593.
- [19] A.J. Ebele, M. Abou-Elwafa Abdallah, S. Harrad, Pharmaceuticals and personal care products (PPCPs) in the freshwater aquatic environment, Emerging Contaminants, 3 (2017) 1-16 https://doi.org/10.1016/j.emcon.2016.12.004.
- [20] M.J. Benotti, R.A. Trenholm, B.J. Vanderford, J.C. Holady, B.D. Stanford, S.A. Snyder, Pharmaceuticals and Endocrine Disrupting Compounds in U.S. Drinking Water, Environmental Science & Technology, 43 (2009) 597-603 https://doi.org/10.1021/es801845a.
- [21] G.M. Bruce, R.C. Pleus, S.A. Snyder, Toxicological Relevance of Pharmaceuticals in Drinking Water, Environmental Science & Technology, 44 (2010) 5619-5626 https://doi.org/10.1021/es1004895.

- [22] S. Nayak, B.P. Chaplin, Fabrication and characterization of porous, conductive, monolithic Ti4O7 electrodes, Electrochimica Acta, 263 (2018) 299-310 https://doi.org/10.1016/j.electacta.2018.01.034.
- [23] M. Najafpour, N. Jameei Moghaddam, S. Hosseini, S. Madadkhani, M. Hołyńska, S. Mehrabani, R. Bagheri, Z. Song, Nanolayered manganese oxide: Insights from inorganic electrochemistry, Catalysis Science & Technology, 7 (2017) 176-185 https://doi.org/10.1039/C7CY00215G.
- [24] Y. Jing, B.P. Chaplin, Mechanistic Study of the Validity of Using Hydroxyl Radical Probes To Characterize Electrochemical Advanced Oxidation Processes, Environmental Science & Technology, 51 (2017) 2355-2365 https://doi.org/10.1021/acs.est.6b05513.
- [25] E. Cuervo Lumbaque, L. Baptista-Pires, J. Radjenovic, Functionalization of graphene sponge electrodes with two-dimensional materials for tailored electrocatalytic activity towards specific contaminants of emerging concern, Chemical Engineering Journal, 446 (2022) 137057 https://doi.org/10.1016/j.cej.2022.137057.
- [26] A.N. Baga, G.R.A. Johnson, N.B. Nazhat, R.A. Saadalla-Nazhat, A simple spectrophotometric determination of hydrogen peroxide at low concentrations in aqueous solution, Analytica Chimica Acta, 204 (1988) 349-353 https://doi.org/10.1016/S0003-2670(00)86374-6.
- [27] P. Roy, S. Berger, P. Schmuki, TiO2 Nanotubes: Synthesis and Applications, Angewandte Chemie International Edition, 50 (2011) 2904-2939 https://doi.org/10.1002/anie.201001374.
- [28] M.C. Biesinger, L.W.M. Lau, A.R. Gerson, R.S.C. Smart, Resolving surface chemical states in XPS analysis of first row transition metals, oxides and hydroxides: Sc, Ti, V, Cu and Zn, Applied Surface Science, 257 (2010) 887-898 https://doi.org/10.1016/j.apsusc.2010.07.086.
- [29] K.A. Soliman, A.F. Zedan, A. Khalifa, H.A. El-Sayed, A.S. Aljaber, S.Y. AlQaradawi, N.K. Allam, Silver Nanoparticles-Decorated Titanium Oxynitride Nanotube Arrays for Enhanced Solar Fuel Generation, Scientific Reports, 7 (2017) 1913 https://doi.org/10.1038/s41598-017-02124-1.
- [30] K. Ohwada, On the pauling electronegativity scales—II, Polyhedron, 3 (1984) 853-859 https://doi.org/10.1016/S0277-5387(00)84634-3.
- [31] C.C.L. McCrory, S. Jung, J.C. Peters, T.F. Jaramillo, Benchmarking Heterogeneous Electrocatalysts for the Oxygen Evolution Reaction, Journal of the American Chemical Society, 135 (2013) 16977-16987 https://doi.org/10.1021/ja407115p.
- [32] C.-H. Chen, E.C. Njagi, S.-Y. Chen, D.T. Horvath, L. Xu, A. Morey, C. Mackin, R. Joesten, S.L. Suib, Structural Distortion of Molybdenum-Doped Manganese Oxide Octahedral Molecular Sieves for Enhanced Catalytic Performance, Inorganic Chemistry, 54 (2015) 10163-10171 https://doi.org/10.1021/acs.inorgchem.5b00906.
- [33] P. Hosseini-Benhangi, C.H. Kung, A. Alfantazi, E.L. Gyenge, Controlling the Interfacial Environment in the Electrosynthesis of MnOx Nanostructures for High-Performance Oxygen Reduction/Evolution Electrocatalysis, ACS Applied Materials & Interfaces, 9 (2017) 26771-26785 https://doi.org/10.1021/acsami.7b05501.
- [34] B. Babakhani, D.G. Ivey, Effect of electrodeposition conditions on the electrochemical capacitive behavior of synthesized manganese oxide electrodes, Journal of Power Sources, 196 (2011) 10762-10774 https://doi.org/10.1016/j.jpowsour.2011.08.102.
- [35] M. Nakayama, A. Tanaka, Y. Sato, T. Tonosaki, K. Ogura, Electrodeposition of Manganese and Molybdenum Mixed Oxide Thin Films and Their Charge Storage Properties, Langmuir, 21 (2005) 5907-5913 https://doi.org/10.1021/la050114u.
- [36] S.E. Balaghi, C.A. Triana, G.R. Patzke, Molybdenum-Doped Manganese Oxide as a Highly Efficient and Economical Water Oxidation Catalyst, ACS Catalysis, 10 (2020) 2074-2087 https://doi.org/10.1021/acscatal.9b02718.

- [37] Y. Zhang, H. Zhu, U. Szewzyk, S. Lübbecke, S. Uwe Geissen, Removal of emerging organic contaminants with a pilot-scale biofilter packed with natural manganese oxides, Chemical Engineering Journal, 317 (2017) 454-460 https://doi.org/10.1016/j.cej.2017.02.095.
- [38] I. Forrez, M. Carballa, K. Verbeken, L. Vanhaecke, M. Schlüsener, T. Ternes, N. Boon, W. Verstraete, Diclofenac Oxidation by Biogenic Manganese Oxides, Environmental Science & Technology, 44 (2010) 3449-3454 https://doi.org/10.1021/es9027327.
- [39] H. Zhang, C.-H. Huang, Oxidative Transformation of Triclosan and Chlorophene by Manganese Oxides, Environmental Science & Technology, 37 (2003) 2421-2430 https://doi.org/10.1021/es026190q.
- [40] H. Tamura, T. Oda, M. Nagayama, R. Furuichi, Acid-Base Dissociation of Surface Hydroxyl Groups on Manganese Dioxide in Aqueous Solutions, Journal of The Electrochemical Society, 136 (1989) 2782-2786 https://doi.org/10.1149/1.2096286.
- [41] A.T. Stone, J.J. Morgan, Reduction and dissolution of manganese(III) and manganese(IV) oxides by organics: 2. Survey of the reactivity of organics, Environmental Science & Technology, 18 (1984) 617-624 https://doi.org/10.1021/es00126a010.
- [42] M. Huguet, M. Deborde, S. Papot, H. Gallard, Oxidative decarboxylation of diclofenac by manganese oxide bed filter, Water Research, 47 (2013) 5400-5408 https://doi.org/10.1016/j.watres.2013.06.016.
- [43] N. Sergienko, J. Radjenovic, Manganese oxide coated TiO2 nanotube-based electrode for efficient and selective electrocatalytic sulfide oxidation to colloidal sulfur, Applied Catalysis B: Environmental, 296 (2021) 120-383 https://doi.org/10.1016/j.apcatb.2021.120383.
- [44] R. Eppstein, M. Caspary Toroker, On the Interplay Between Oxygen Vacancies and Small Polarons in Manganese Iron Spinel Oxides, ACS Materials Au, 2 (2022) 269-277 https://doi.org/10.1021/acsmaterialsau.1c00051.
- [45] J. Zeng, H. Xie, Z. Liu, X. Liu, G. Zhou, Y. Jiang, Oxygen vacancy induced MnO2 catalysts for efficient toluene catalytic oxidation, Catalysis Science & Technology, 11 (2021) 6708-6723 10.1039/D1CY01274F.
- [46] G.A. Kimmel, N.G. Petrik, Tetraoxygen on reduced TiO2(110): oxygen adsorption and reactions with bridging oxygen vacancies, Physical review letters, 100 (2008) 196102 https://doi.org/10.1103/PhysRevLett.100.196102.
- [47] J. Chen, W. Chen, M. Huang, H. Tang, J. Zhang, G. Wang, R. Wang, Metal organic frameworks derived manganese dioxide catalyst with abundant chemisorbed oxygen and defects for the efficient removal of gaseous formaldehyde at room temperature, Applied Surface Science, 565 (2021) 150445 https://doi.org/10.1016/j.apsusc.2021.150445.
- [48] M.M. Montemore, M.A. van Spronsen, R.J. Madix, C.M. Friend, O2 Activation by Metal Surfaces: Implications for Bonding and Reactivity on Heterogeneous Catalysts, Chemical Reviews, 118 (2018) 2816-2862 https://doi.org/10.1021/acs.chemrev.7b00217.
- [49] L. Geng, B. Chen, J. Yang, C. Shui, Y. Songshou, F. Jile, N. Zhang, J. Xie, B. Chen, Synergistic effect between Mn and Ce for active and stable catalytic wet air oxidation of phenol over MnCeOx, Applied Catalysis A: General, 604 (2020) 117-774 https://doi.org/10.1016/j.apcata.2020.117774.
- [50] L.-F. Zhai, Z.-X. Chen, J.-X. Qi, M. Sun, Manganese-doped molybdenum oxide boosts catalytic performance of electrocatalytic wet air oxidation at ambient temperature, Journal of Hazardous Materials, 428 (2022) 128-245 https://doi.org/10.1016/j.jhazmat.2022.128245.
- [51] A. Massa, S. Hernández, A. Lamberti, C. Galletti, N. Russo, D. Fino, Electro-oxidation of phenol over electrodeposited MnOx nanostructures and the role of a TiO2 nanotubes interlayer, Applied Catalysis B: Environmental, 203 (2017) 270-281 https://doi.org/10.1016/j.apcatb.2016.10.025.

- [52] Y. Jiang, H. Zhao, J. Liang, L. Yue, T. Li, Y. Luo, Q. Liu, S. Lu, A.M. Asiri, Z. Gong, X. Sun, Anodic oxidation for the degradation of organic pollutants: Anode materials, operating conditions and mechanisms. A mini review, Electrochemistry Communications, 123 (2021) 106912 https://doi.org/10.1016/j.elecom.2020.106912.
- [53] L. Miao, W. Deng, X. Chen, M. Gao, W. Chen, T. Ao, Selective adsorption of phosphate by carboxyl-modified activated carbon electrodes for capacitive deionization, Water science and technology : a journal of the International Association on Water Pollution Research, 84 (2021) 1757-1773 https://doi.org/10.2166/wst.2021.358.
- [54] M. Soufan, M. Deborde, B. Legube, Aqueous chlorination of diclofenac: kinetic study and transformation products identification, Water Research, 46 (2012) 3377-3386 https://doi.org/10.1016/j.watres.2012.03.056.
- [55] H. Cheng, D. Song, Y. Chang, H. Liu, J. Qu, Chlorination of tramadol: Reaction kinetics, mechanism and genotoxicity evaluation, Chemosphere, 141 (2015) 282-289 https://doi.org/10.1016/j.chemosphere.2015.06.034.
- [56] R. Xie, X. Meng, P. Sun, J. Niu, W. Jiang, L. Bottomley, D. Li, Y. Chen, J. Crittenden, Electrochemical oxidation of ofloxacin using a TiO2-based SnO2-Sb/polytetrafluoroethylene resin-PbO2 electrode: Reaction kinetics and mass transfer impact, Applied Catalysis B: Environmental, 203 (2017) 515-525 https://doi.org/10.1016/j.apcatb.2016.10.057.
- [57] N. Ormeno-Cano, J. Radjenovic, Electrochemical degradation of antibiotics using flow-through graphene sponge electrodes, Journal of Hazardous Materials, 431 (2022) 128-462 https://doi.org/10.1016/j.jhazmat.2022.128462.
- [58] S. Franz, E. Falletta, H. Arab, S. Murgolo, M. Bestetti, G. Mascolo, Degradation of Carbamazepine by Photo(electro)catalysis on Nanostructured TiO2 Meshes: Transformation Products and Reaction Pathways, Catalysts, 2020. https://doi.org/10.3390/catal10020169
- [59] Y. Pan, S. Cheng, X. Yang, J. Ren, J. Fang, C. Shang, W. Song, L. Lian, X. Zhang, UV/chlorine treatment of carbamazepine: Transformation products and their formation kinetics, Water Research, 116 (2017) 254-265 https://doi.org/10.1016/j.watres.2017.03.033.
- [60] S. Cravanzola, M. Sarro, F. Cesano, P. Calza, D. Scarano, Few-Layer MoS2 Nanodomains Decorating TiO2 Nanoparticles: A Case Study for the Photodegradation of Carbamazepine, Nanomaterials, 2018. https://doi.org/10.1016/j.watres.10.3390/nano8040207
- [61] J. Zhai, Q. Wang, Q. Li, B. Shang, M.H. Rahaman, J. Liang, J. Ji, W. Liu, Degradation mechanisms of carbamazepine by δ-MnO2: Role of protonation of degradation intermediates, Science of The Total Environment, 640-641 (2018) 981-988 https://doi.org/10.1016/j.scitotenv.2018.05.368.
- [62] T. Kosjek, H.R. Andersen, B. Kompare, A. Ledin, E. Heath, Fate of Carbamazepine during Water Treatment, Environmental Science & Technology, 43 (2009) 6256-6261 https://doi.org/10.1021/es900070h.
- [63] A. Jelic, I. Michael, A. Achilleos, E. Hapeshi, D. Lambropoulou, S. Perez, M. Petrovic, D. Fatta-Kassinos, D. Barcelo, Transformation products and reaction pathways of carbamazepine during photocatalytic and sonophotocatalytic treatment, Journal of Hazardous Materials, 263 (2013) 177-186 https://doi.org/10.1016/j.jhazmat.2013.07.068.
- [64] J. Li, L. Dodgen, Q. Ye, J. Gan, Degradation Kinetics and Metabolites of Carbamazepine in Soil, Environmental Science & Technology, 47 (2013) 3678-3684 https://doi.org/10.1021/es304944c.
- [65] L. Hu, H.M. Martin, O. Arce-Bulted, M.N. Sugihara, K.A. Keating, T.J. Strathmann, Oxidation of Carbamazepine by Mn(VII) and Fe(VI): Reaction Kinetics and Mechanism, Environmental Science & Technology, 43 (2009) 509-515 https://doi.org/10.1021/es8023513.
- [66] Y. Ding, G. Zhang, X. Wang, L. Zhu, H. Tang, Chemical and photocatalytic oxidative degradation of carbamazepine by using metastable Bi3+ self-doped NaBiO3 nanosheets as a bifunctional

material, Applied Catalysis B: Environmental, 202 (2017) 528-538 https://doi.org/10.1016/j.apcatb.2016.09.054.

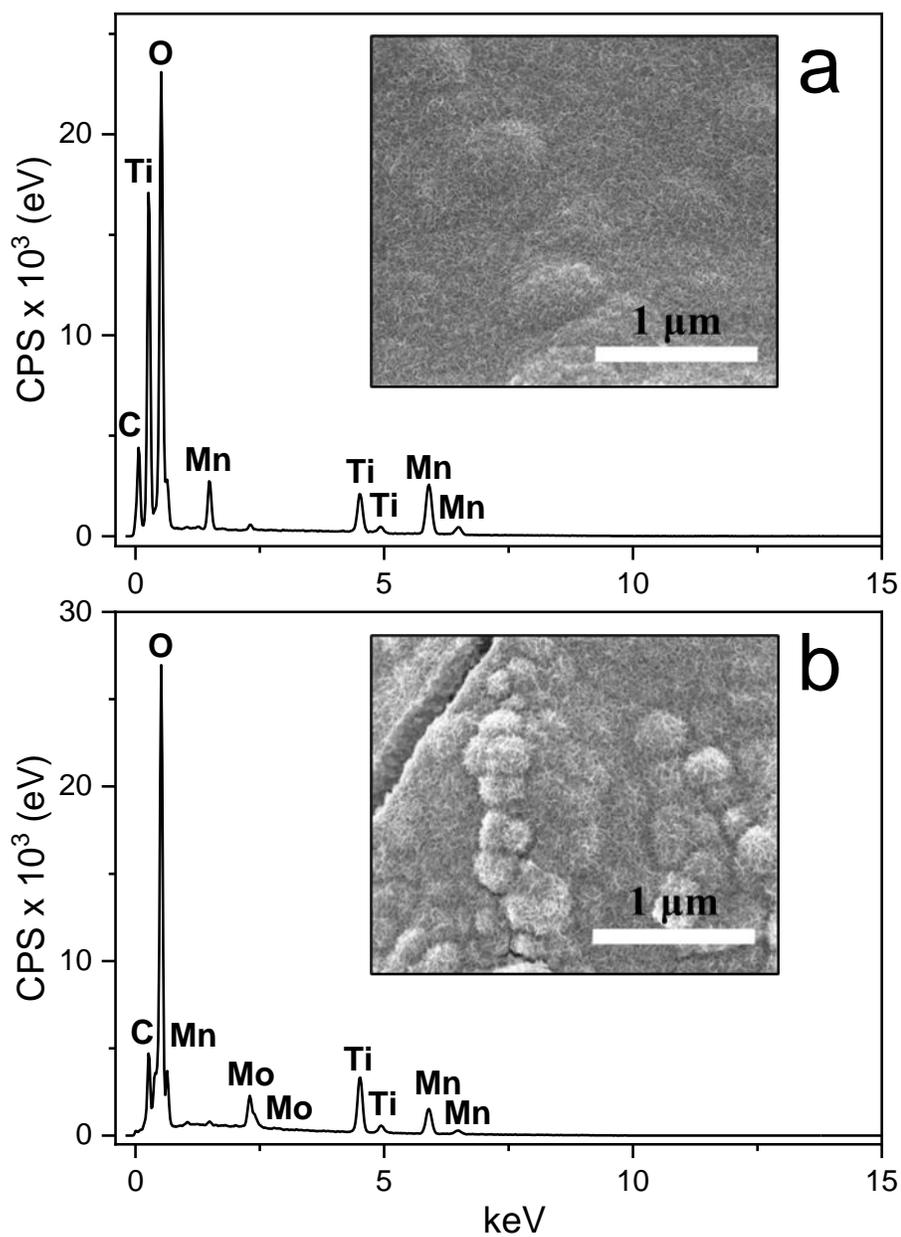

**Figure 1**. FESEM images and EDX spectra of the **a)** Ti/TiO$_2$ NTA–Mn$_x$O$_y$, **b)** Ti/TiO$_2$ NTA–Mn$_x$O$_y$-Mo, synthesized in the bath containing MnO$_4^{2-}$ and Mn$^{2+}$ in the ratio of 1:1000.

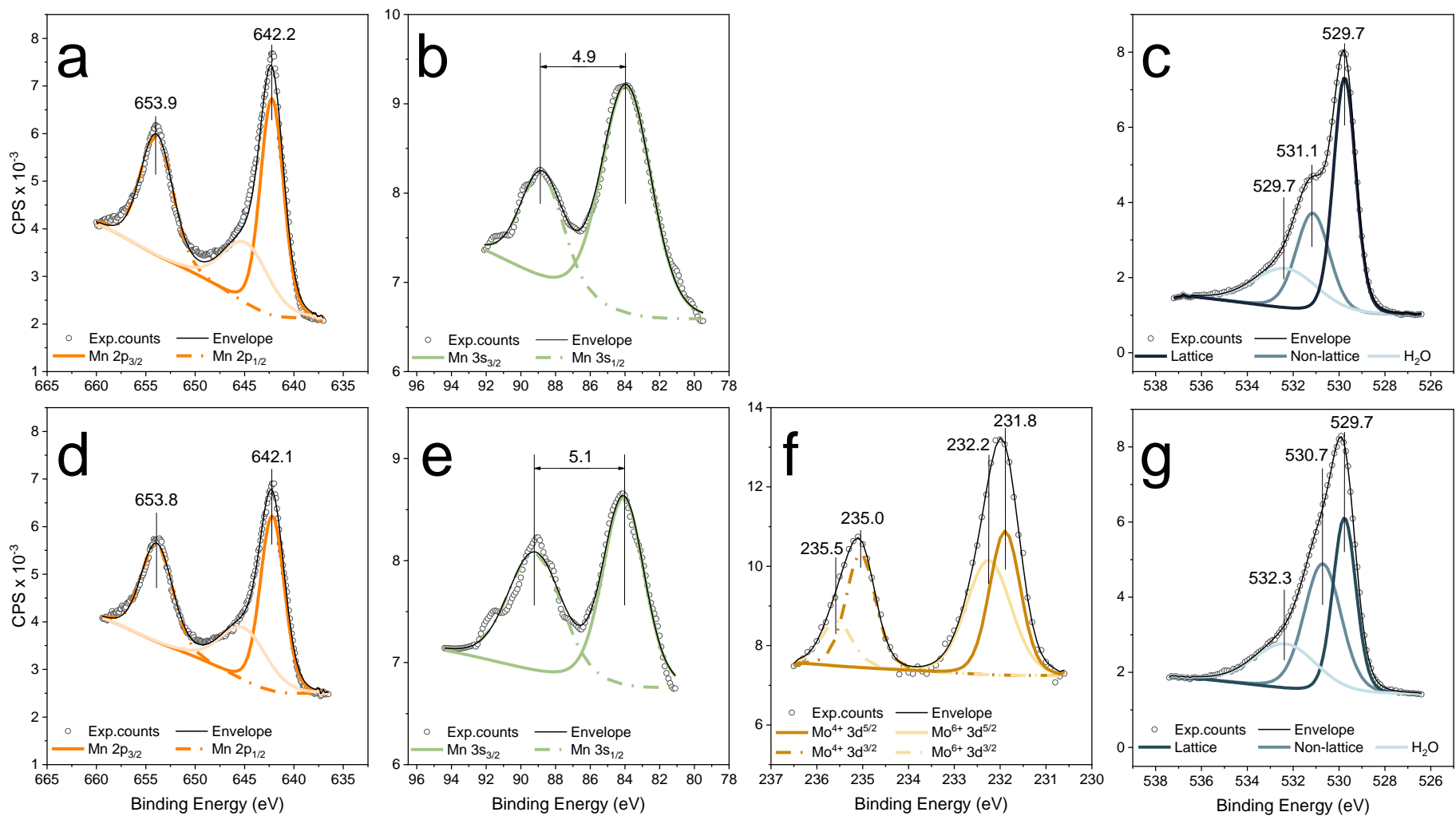

**Figure 2.** XPS spectra displaying the **a)** Mn 2p binding energy, **b)** Mn 3s binding energy, **c)** O1s binding energy of the Ti/TiO$_2$ NTA–Mn$_x$O$_y$, and **d)** Mn 2p binding energy, **e)** Mn 3s binding energy, **f)** Mo 3d binding energy and **g)** O1s binding energy of Ti/TiO$_2$ NTA–Mn$_x$O$_y$-Mo.

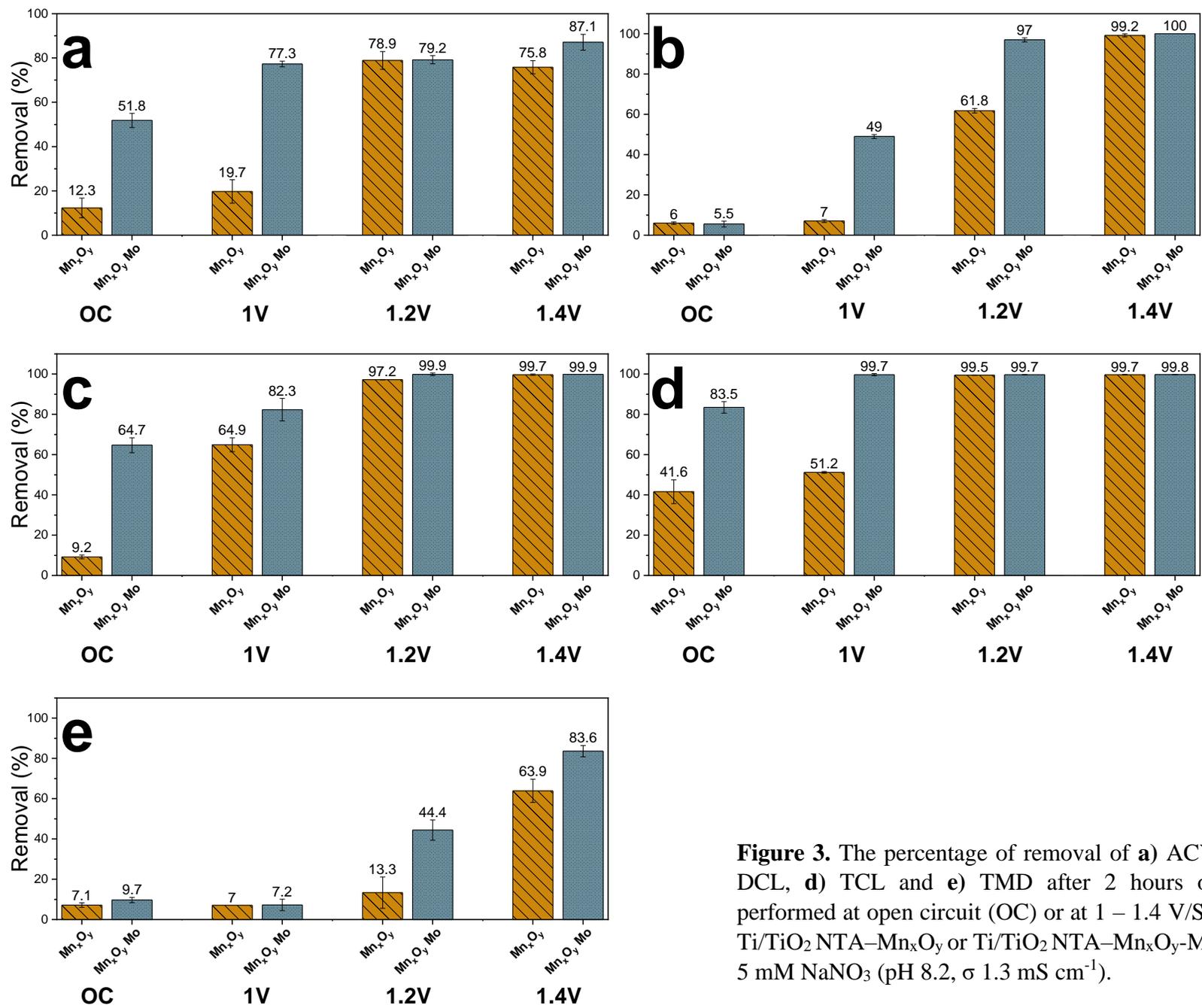

**Figure 3.** The percentage of removal of **a)** ACY, **b)** CBZ, **c)** DCL, **d)** TCL and **e)** TMD after 2 hours of experiment performed at open circuit (OC) or at 1 – 1.4 V/SHE applied to Ti/TiO$_2$ NTA–Mn$_x$O$_y$ or Ti/TiO$_2$ NTA–Mn$_x$O$_y$-Mo anode using 5 mM NaNO$_3$ (pH 8.2, σ 1.3 mS cm$^{-1}$).

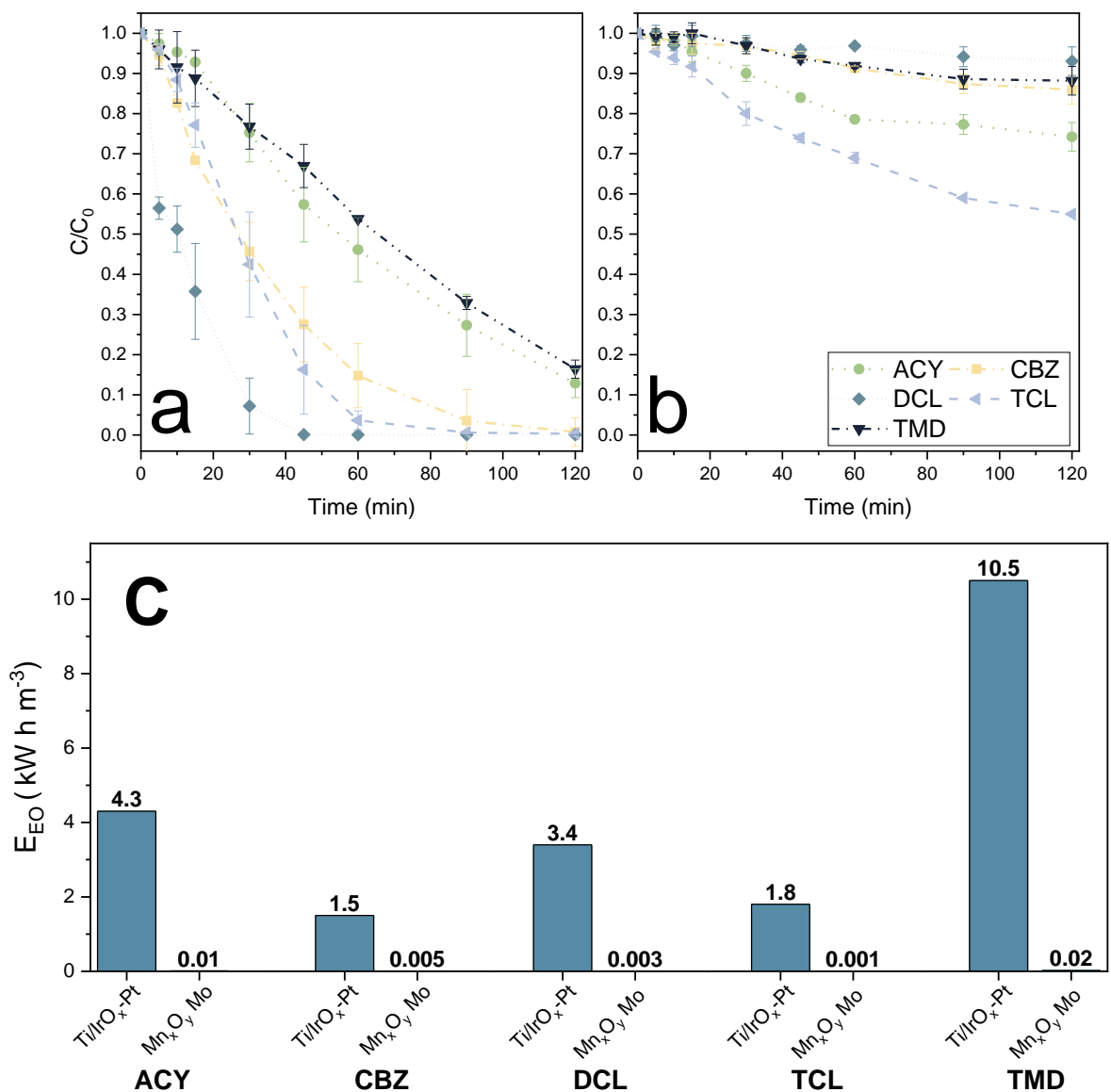

**Figure 4.** Decrease in ACY, CBZ, DCL, TCL, TMD, IPM concentration (C) normalized to the initial value ($C_0$) during experiment performed at 1.4 V/SHE applied to **a)** Ti/TiO$_2$ NTA–Mn$_x$O$_y$-Mo or **b)** Ti/IrO$_x$-Pt anodes, both performed in 5 mM NaNO$_3$ (pH 8.2, σ 1.3 mS cm$^{-1}$) **c)** $E_{EO}$ required for contaminants removal in an electrochemical cell equipped with Ti/TiO$_2$NTA–Mn$_x$O$_y$-Mo, or Ti/IrOx-Pt anodes at 1.4 V/SHE.

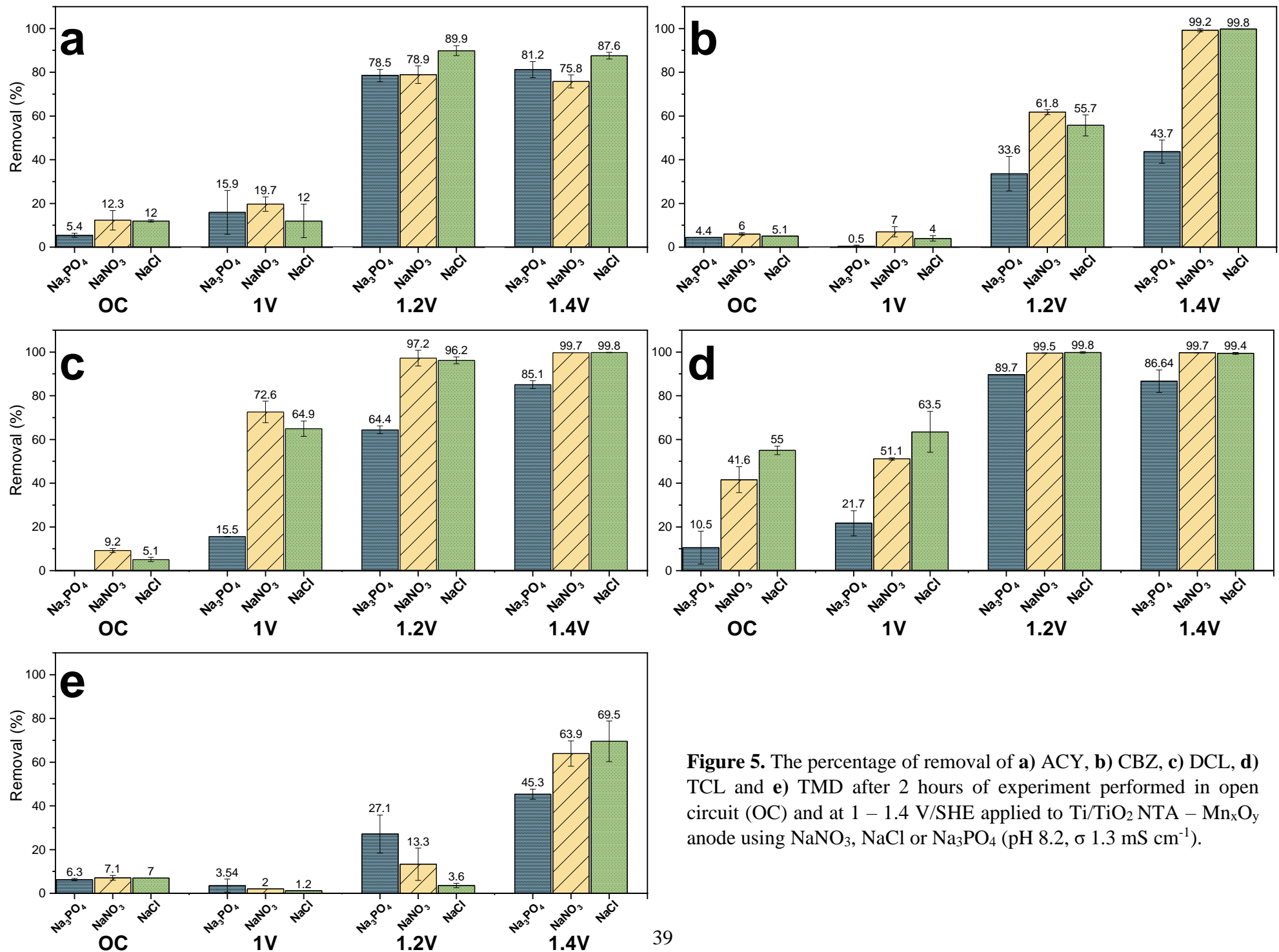

**Figure 5.** The percentage of removal of **a)** ACY, **b)** CBZ, **c)** DCL, **d)** TCL and **e)** TMD after 2 hours of experiment performed in open circuit (OC) and at 1 – 1.4 V/SHE applied to Ti/TiO$_2$ NTA – Mn$_x$O$_y$ anode using NaNO$_3$, NaCl or Na$_3$PO$_4$ (pH 8.2, σ 1.3 mS cm$^{-1}$).

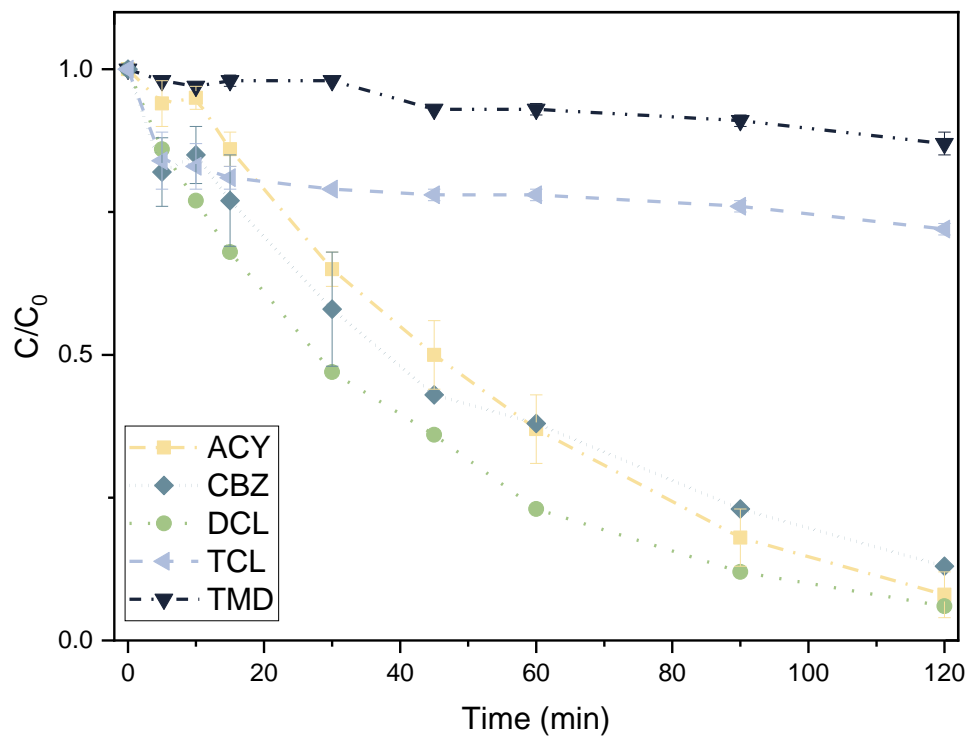

**Figure 6**. Decrease in ACY, CBZ, DCL, TCL, TMD concentration (C) normalized to the initial value ($C_0$) during experiment performed at 1.4 V/SHE applied to Ti/$TiO_2$ NTA – $Mn_xO_y$-Mo anodes performed in tap water (pH 7.5, σ 0.38 mS $cm^{-1}$) purged with $O_2$.

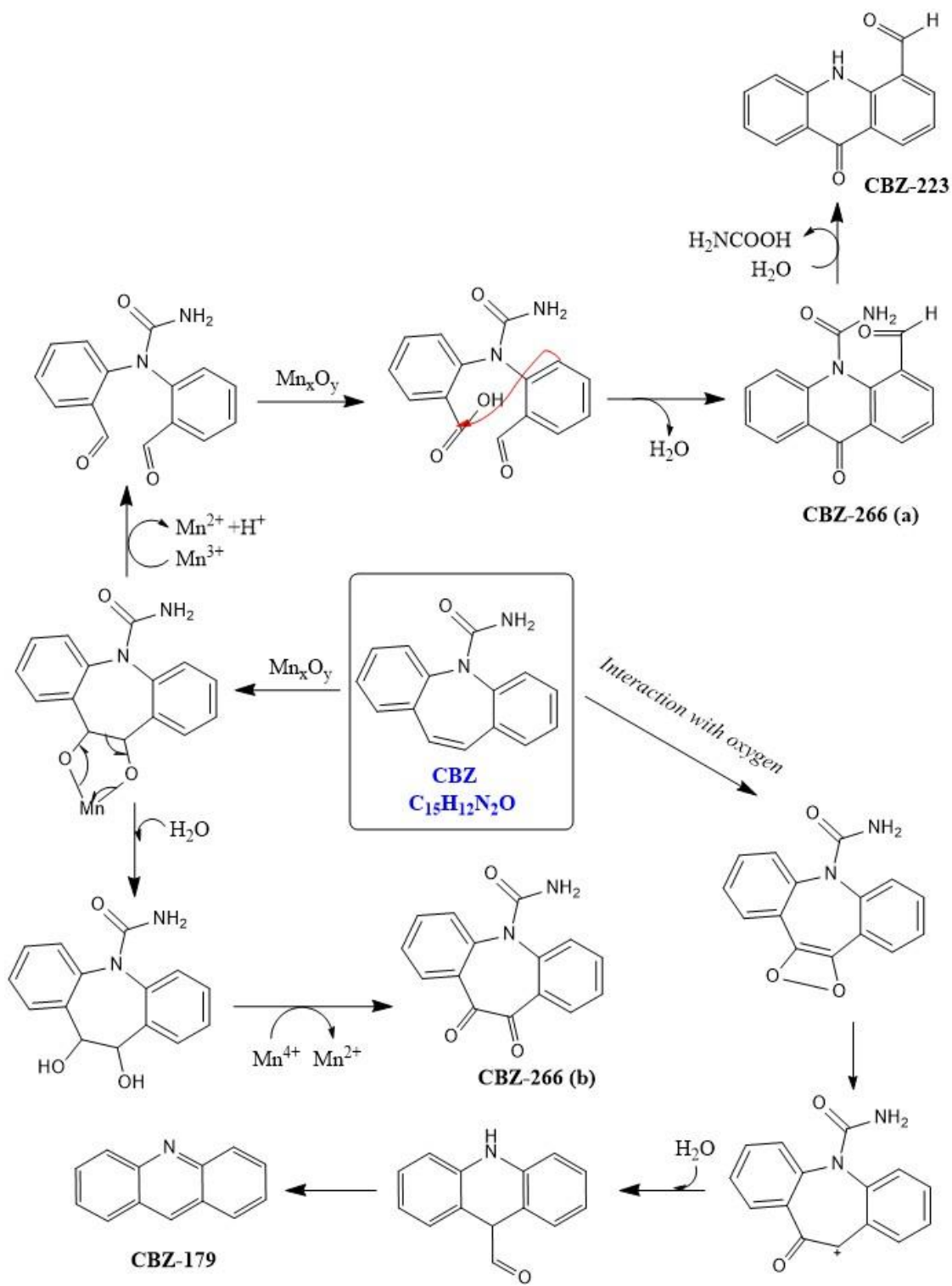

**Figure 7**. Proposed degradation pathway of carbamazepine (CBZ) for electrooxidation at Ti/TiO$_2$NTA – Mn$_x$O$_y$-Mo anode, at 1.4 V/SHE and in 5mM NaNO$_3$ (pH 8.2, σ 1.3 mS cm$^{-1}$).

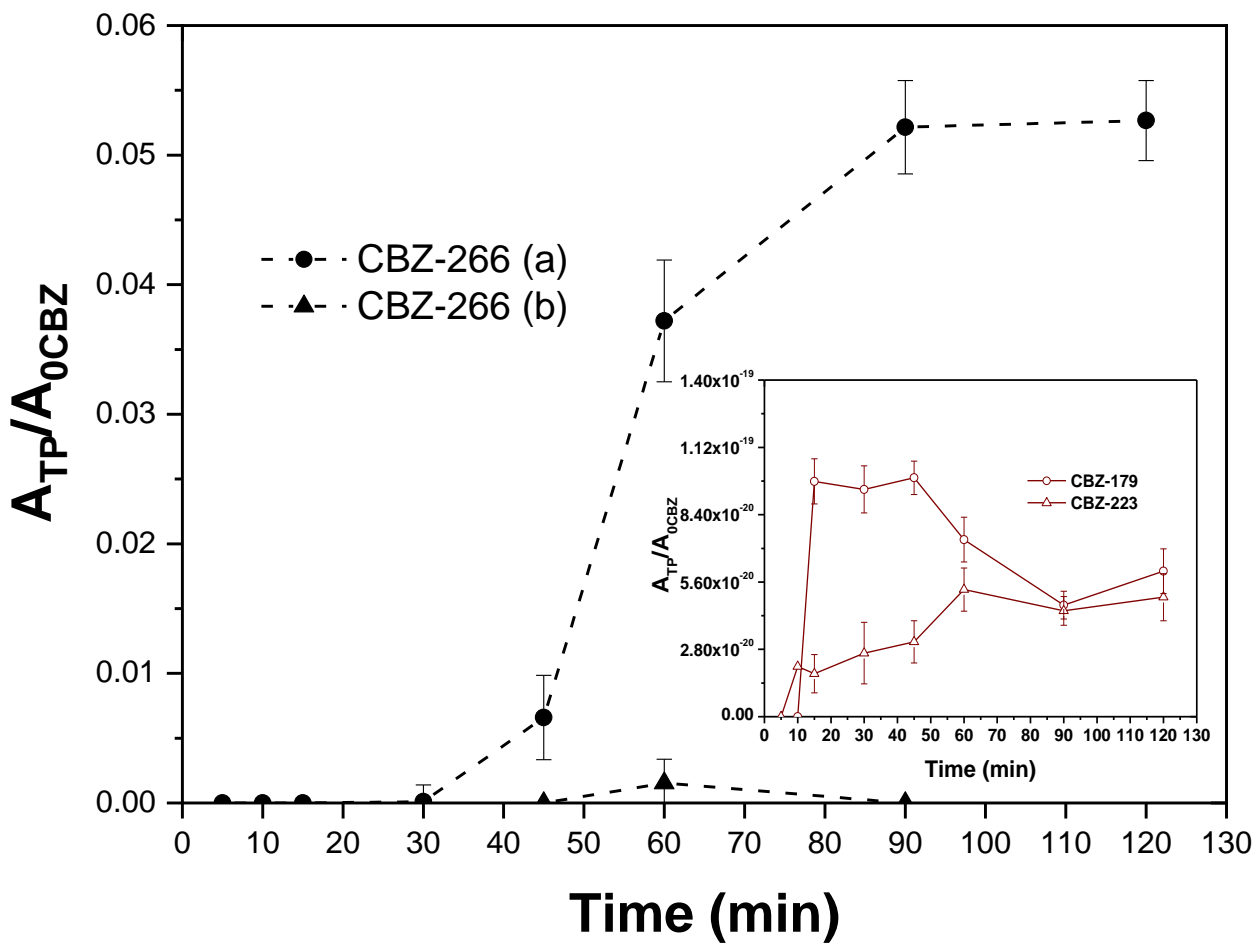

**Figure 8.** Evolution profile of CBZ and its TPs, CBZ-TPs normalized to the initial CBZ area in initial time (A$_0$) during experiment performed at 1.4V/SHE applied to Ti/TiO$_2$ NTA – Mn$_x$O$_y$-Mo anodes performed in 5 mM NaNO$_3$ supporting electrolyte (pH 8.2, σ 1.3 mS cm$^{-1}$).

**Supplementary Material**

**Electrocatalytic removal of persistent organic contaminants at molybdenum doped manganese oxide coated TiO$_2$ nanotube-based anode**

*Natalia Sergienko[a,b], Elisabeth Cuervo Lumbaque[a,b], Nick Duinslaeger[a,b], Jelena Radjenovic[a,c*]*

[a]*Catalan Institute for Water Research (ICRA), Scientific and Technological Park of the University of Girona, 17003 Girona, Spain*

[b] *University of Girona, Girona, Spain*

[c]*Catalan Institution for Research and Advanced Studies (ICREA), Passeig Lluís Companys 23, 08010 Barcelona, Spain*

*\* Corresponding author:*

*Jelena Radjenovic, Catalan Institute for Water Research (ICRA), c/Emili Grahit, 101, 17003 Girona, Spain*

Phone: + 34 972 18 33 80; Fax: +34 972 18 32 48; E-mail: jradjenovic@icra.cat

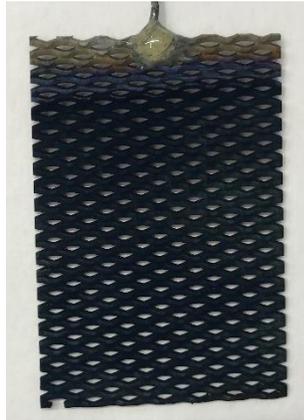

**Figure S1.** The image of synthesized Ti mesh with $TiO_2$ NTA interlayer coated with $Mn_xO_y$.

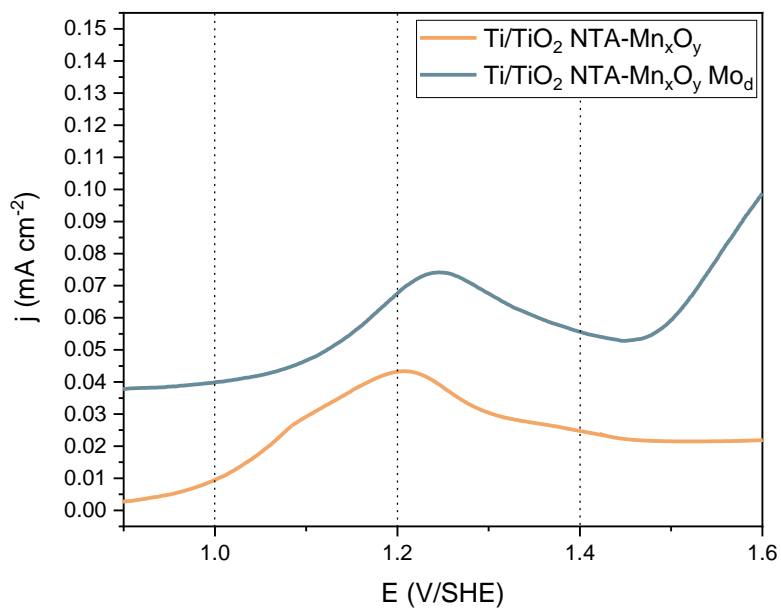

**Figure S2**. Linear sweep voltammetry performed at Ti/TiO$_2$ NTA coated with Mn$_x$O$_y$ and Mo doped materials in 50 mM NaNO$_3$ electrolyte at 20 mV sec$^{-1}$ scan rate.

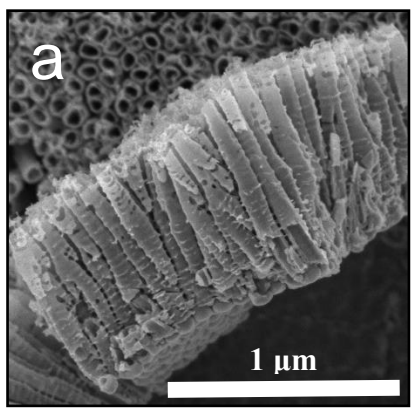
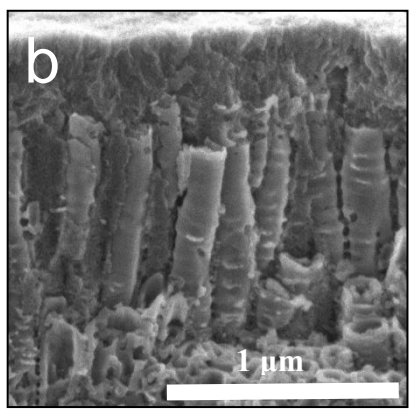
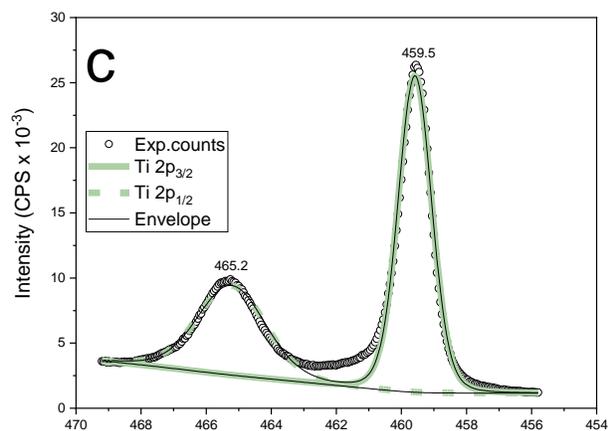
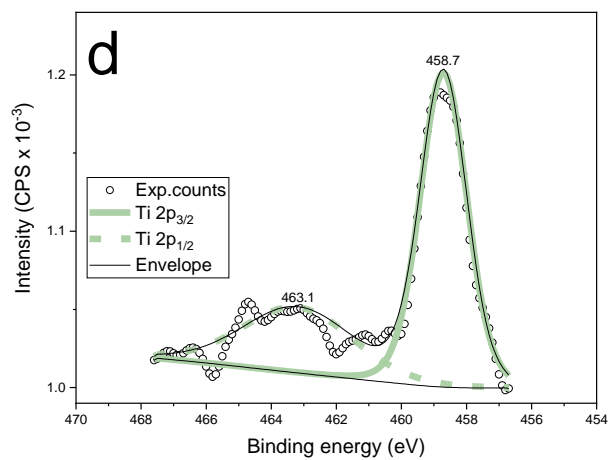

**Figure S3.** FESEM images of the: **a)** cross section of as prepared self-organized TiO$_2$ nanotube array (NTA) and **b)** cross section of self-organized TiO$_2$ filled with Mn$_x$O$_y$. Ti 2p photoelectron spectra for the: **c)** cross section of as prepared self-organized TiO$_2$ NTA and **d)** cross section of self-organized TiO$_2$ filled with Mn$_x$O$_y$.

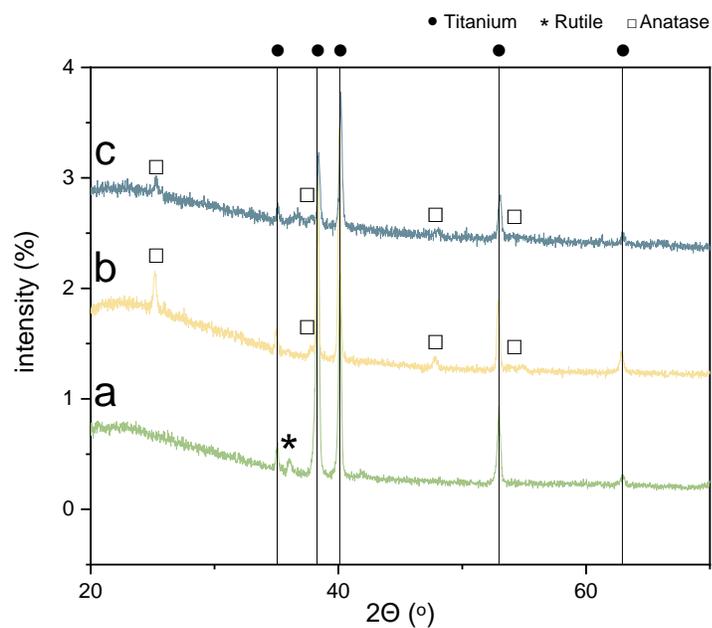

**Figure S4.** The XRD patterns of the: **a)** Ti mesh, **b)** Ti mesh with $TiO_2$ NTA layer after annealing at 400 °C, **c)** Ti/$TiO_2$ NTA coated with $Mn_xO_y$.

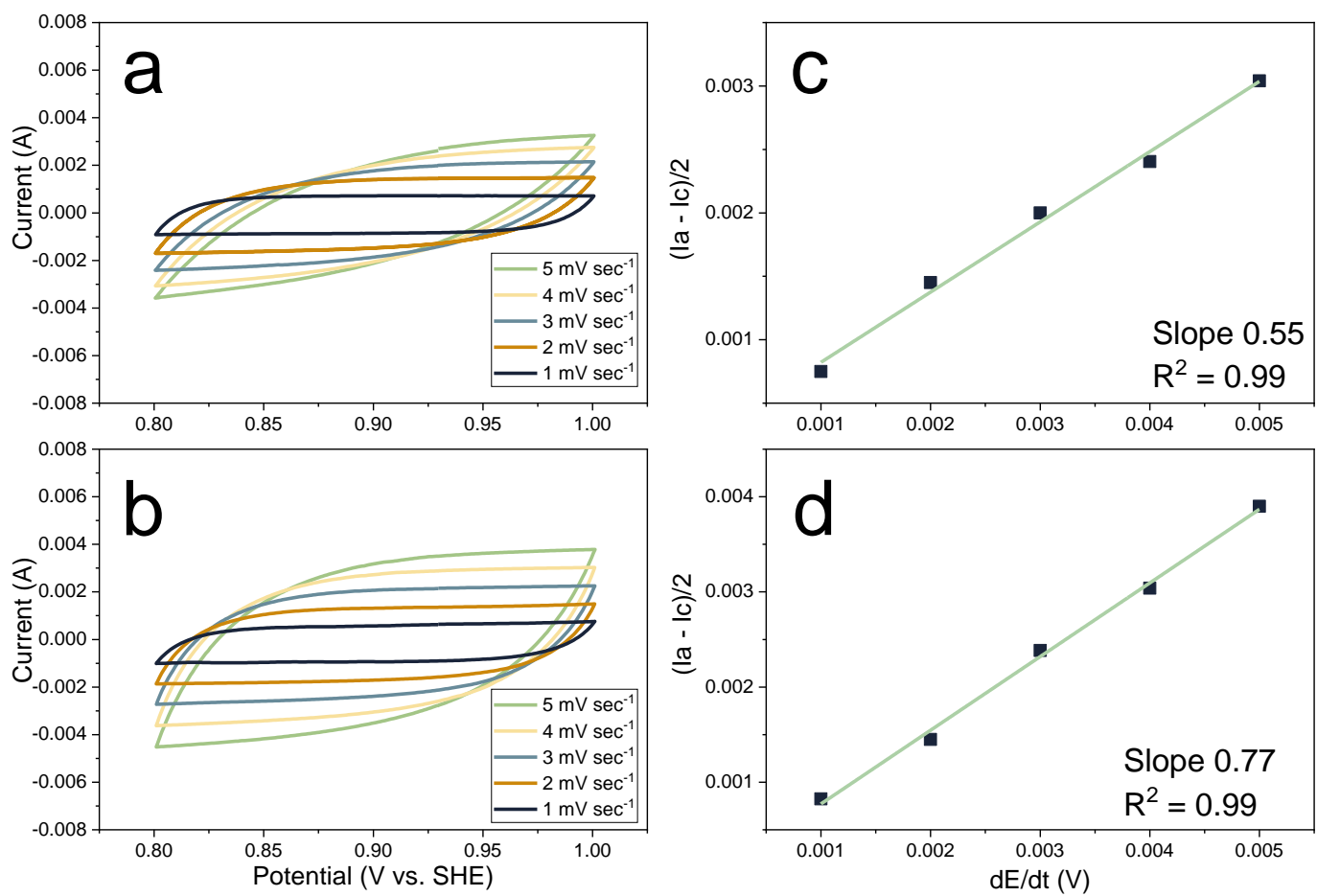

**Figure S5.** Double layer capacitance measured at **a)** Ti/TiO$_2$ NTA coated with Mn$_x$O$_y$, **b)** Ti/TiO$_2$ NTA coated with Mn$_x$O$_y$ doped with Mo and plots of charging currents vs scan rates obtained for **c)** Ti/TiO$_2$ NTA coated with Mn$_x$O$_y$, **d)** Ti/TiO$_2$ NTA coated with Mn$_x$O$_y$ doped with Mo.

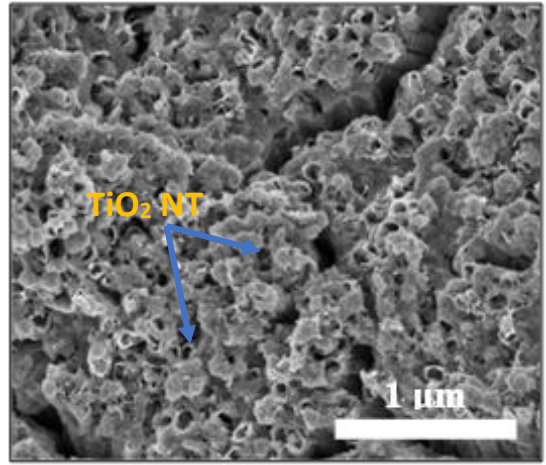

**Figure S6.** FESEM images of the top view of Ti/TiO$_2$ NTA coated with Mn$_x$O$_y$ doped with Mo synthesized in the bath containing MnO$_4^{2-}$ and Mn$^{2+}$ and in the ratio of 1:100.

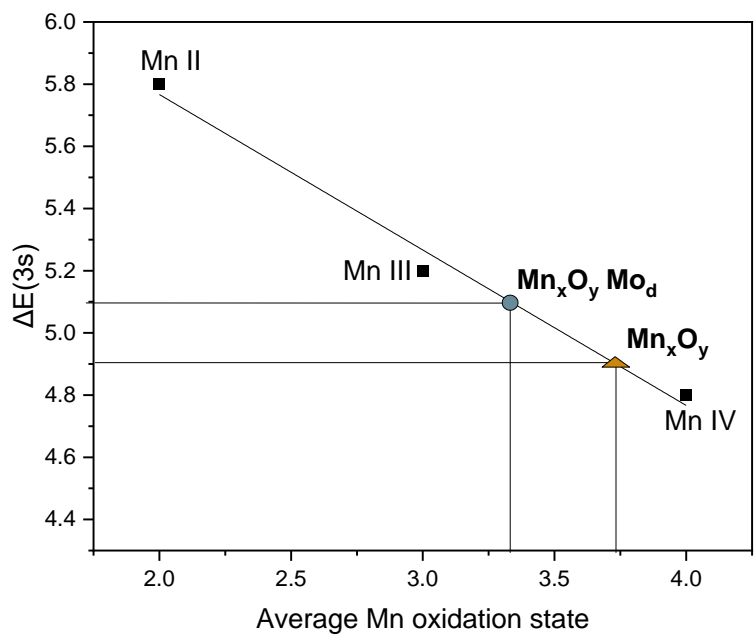

**Figure S7**. XPS linear calibration for quantification of the major Mn oxidation state of Ti/TiO$_2$ NTA – Mn$_x$O$_y$ (brown triangle) and Ti/TiO$_2$ NTA– Mn$_x$O$_y$-Mo (blue circle) based on the ΔE(3s) binding energy of reference compounds Mn II, Mn III, and Mn IV.

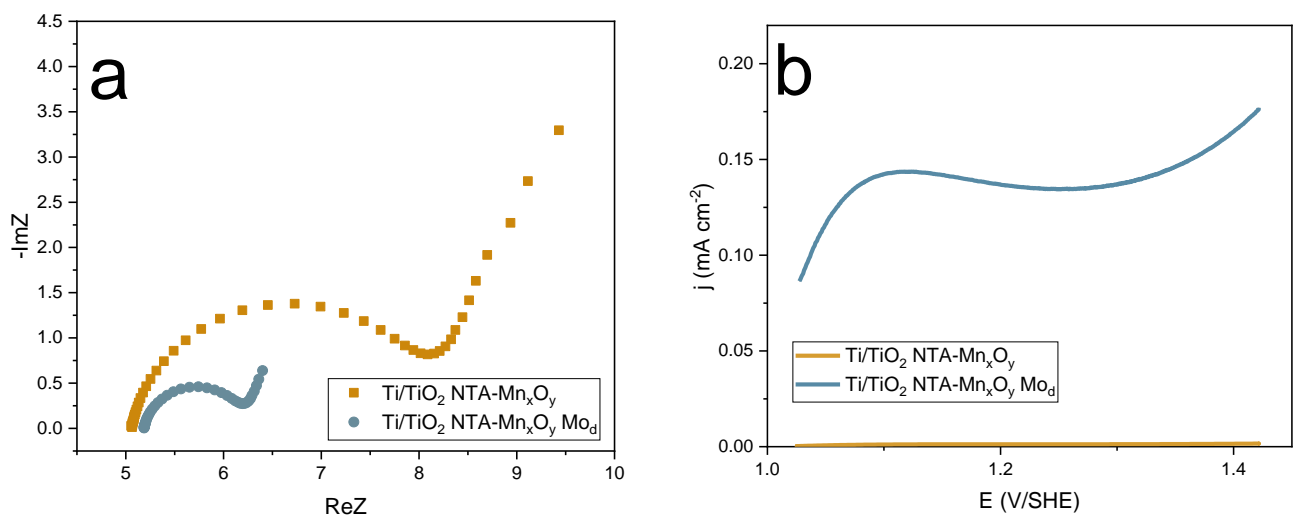

**Figure S8**. **a)** Nyquist plot of Ti/TiO$_2$ NTA coated with Mn$_x$O$_y$ and Mo doped material in 100 mM NaNO$_3$ at 1.4 V/SHE applied potential in the frequency range of 0.1 Hz−50 kHz. **a)** Linear sweep voltammetry performed at Ti/TiO$_2$ NTA coated with Mn$_x$O$_y$ and Mo doped materials in 50 mM NaNO$_3$ electrolyte at 20 mV sec$^{-1}$ scan rate.

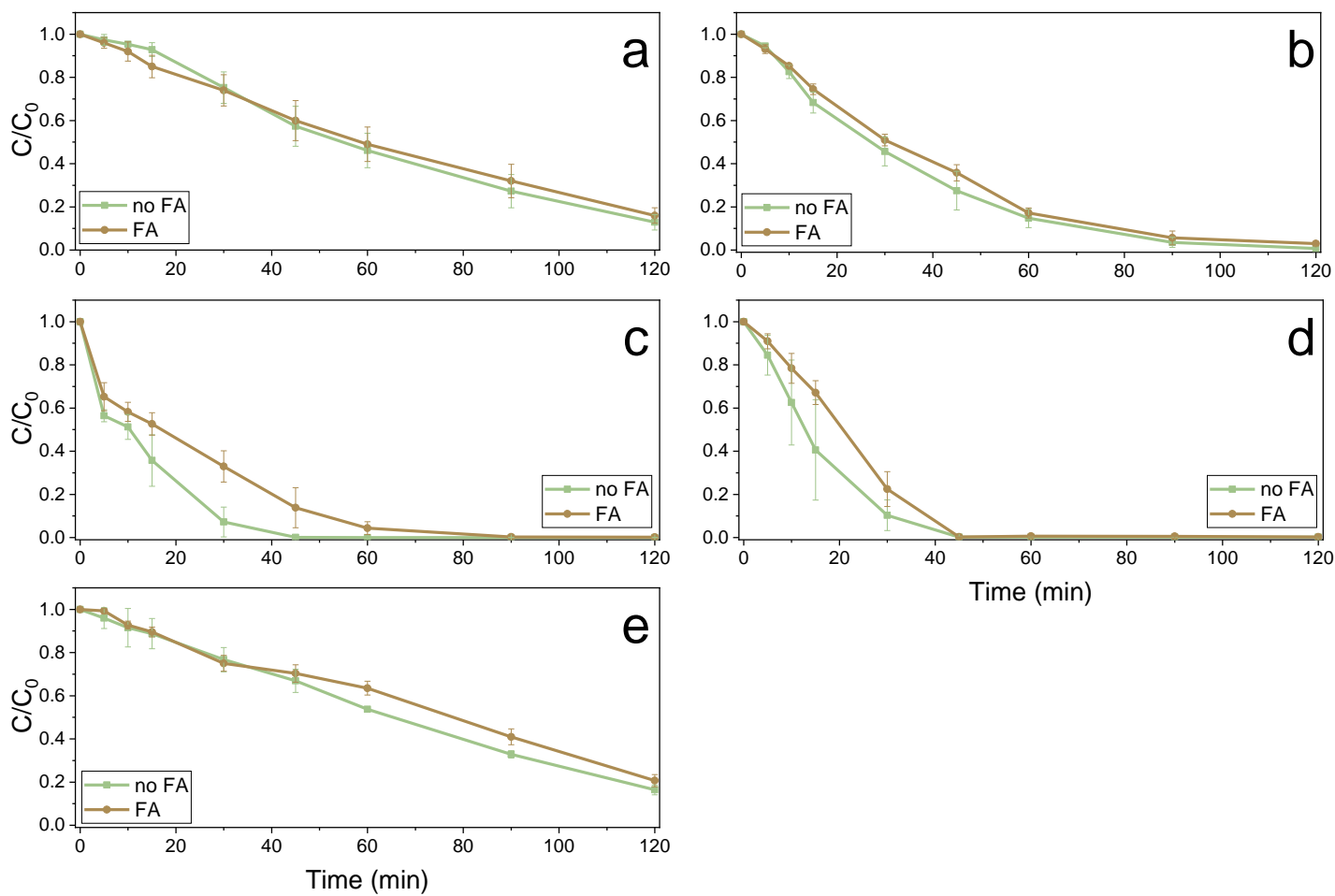

**Figure S9.** Decrease in **a)** ACY, **b)** CBZ, **c)** DCL, **d)** TCL, **e)** TMD concentration (C) normalized to the initial value ($C_0$) during experiment at 1.4 V/SHE applied to Ti/$TiO_2$ NTA – $Mn_xO_y$-Mo anodes performed in 5 mM $NaNO_3$ (pH 8.2, σ 1.3 mS cm$^{-1}$) electrolyte with or without presence of FA.

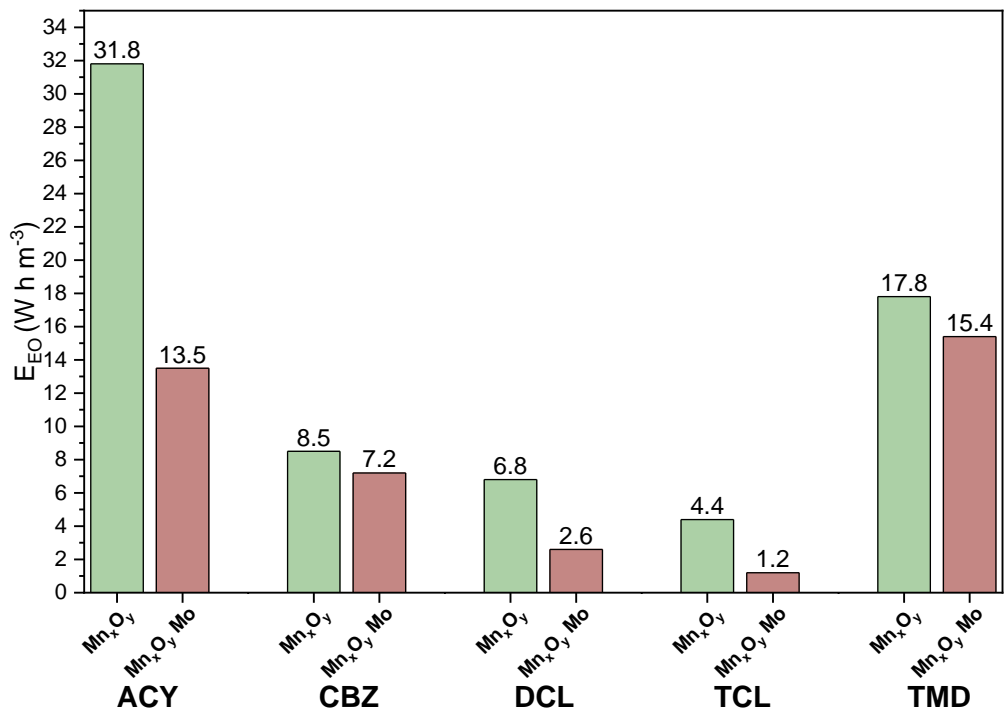

**Figure S10**. $E_{EO}$ (Wh m$^{-3}$) required for contaminants removal in an electrochemical cell equipped with Ti/TiO$_2$ NTA – Mn$_x$O$_y$ or Ti/TiO$_2$ NTA – Mn$_x$O$_y$-Mo anodes at 1.4 V/SHE in 5 mM NaNO$_3$ (pH 8.2, σ 1.3 mS cm$^{-1}$).

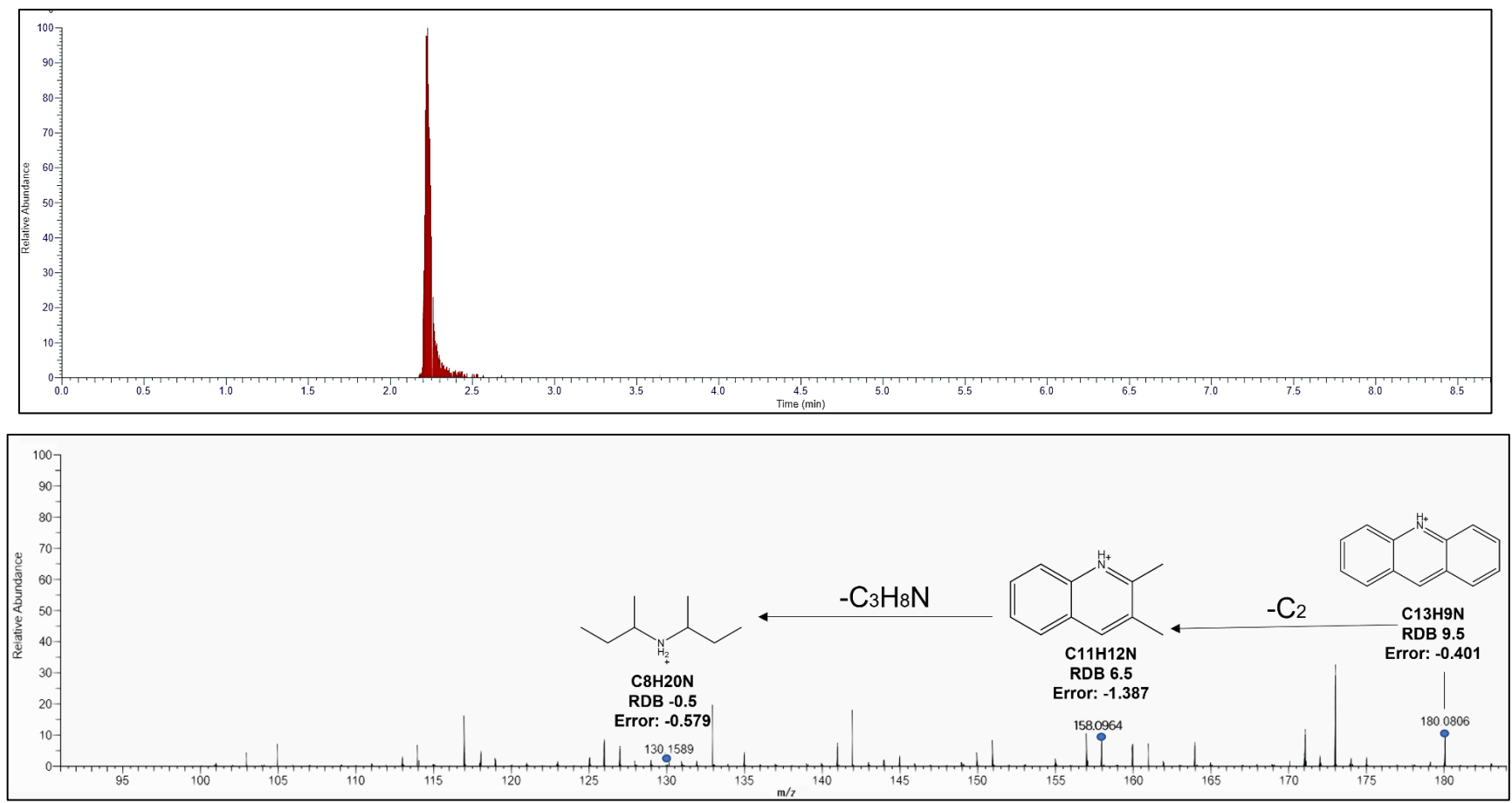

**Figure S11.** MS$^2$ mass spectra of the molecular ion CBZ-179 with proposed structures and losses of fragment ions.

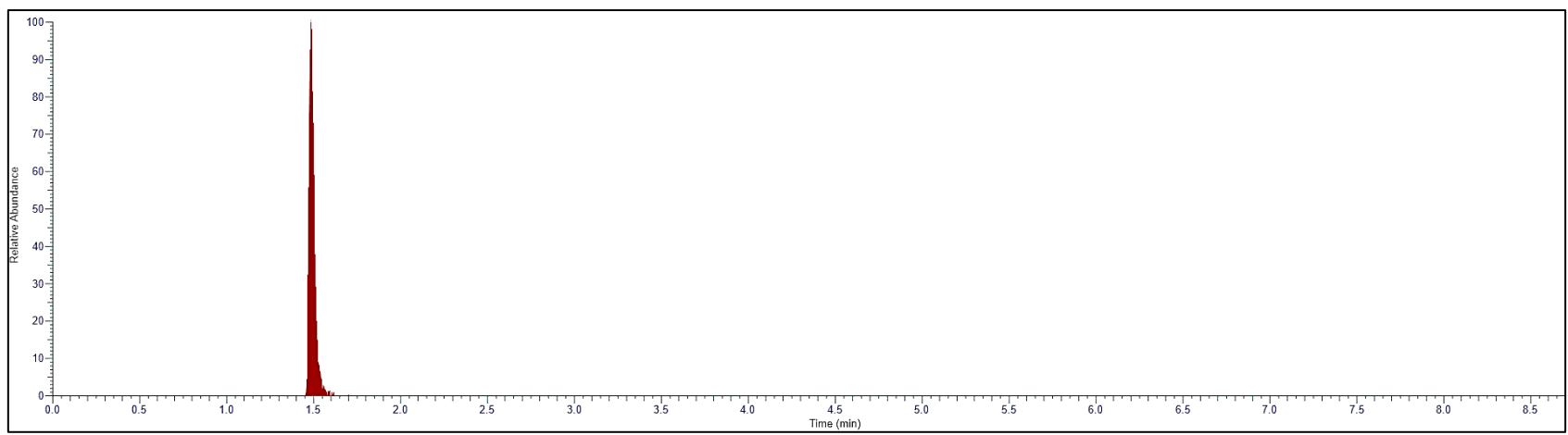
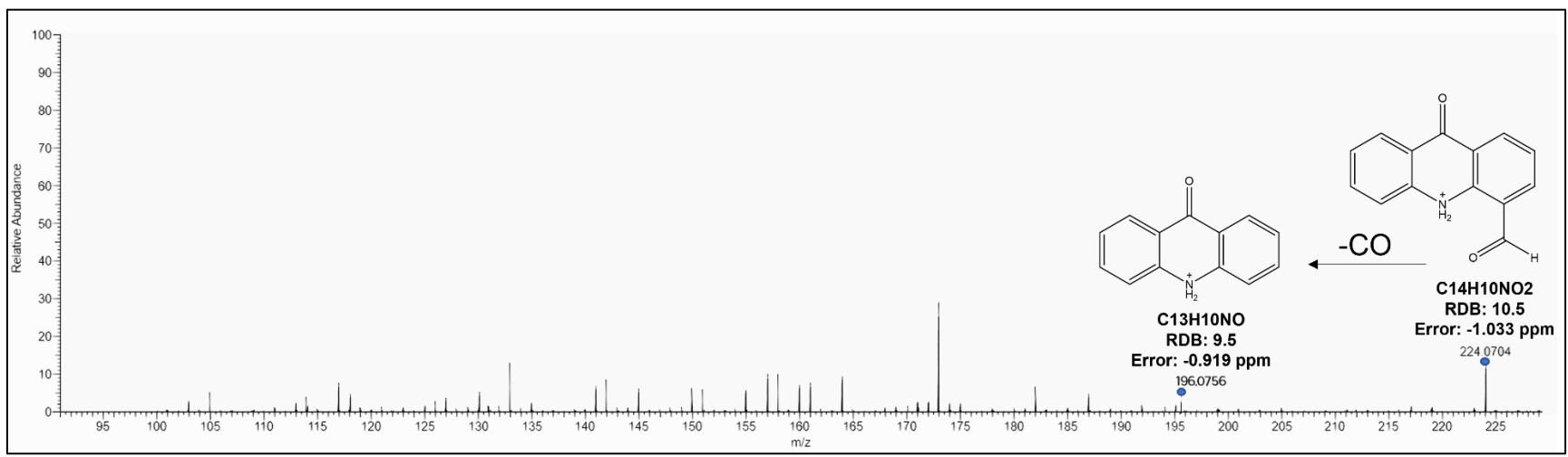

**Figure S12.** MS$^2$ mass spectra of the molecular ion CBZ-223 with proposed structures and losses of fragment ions.

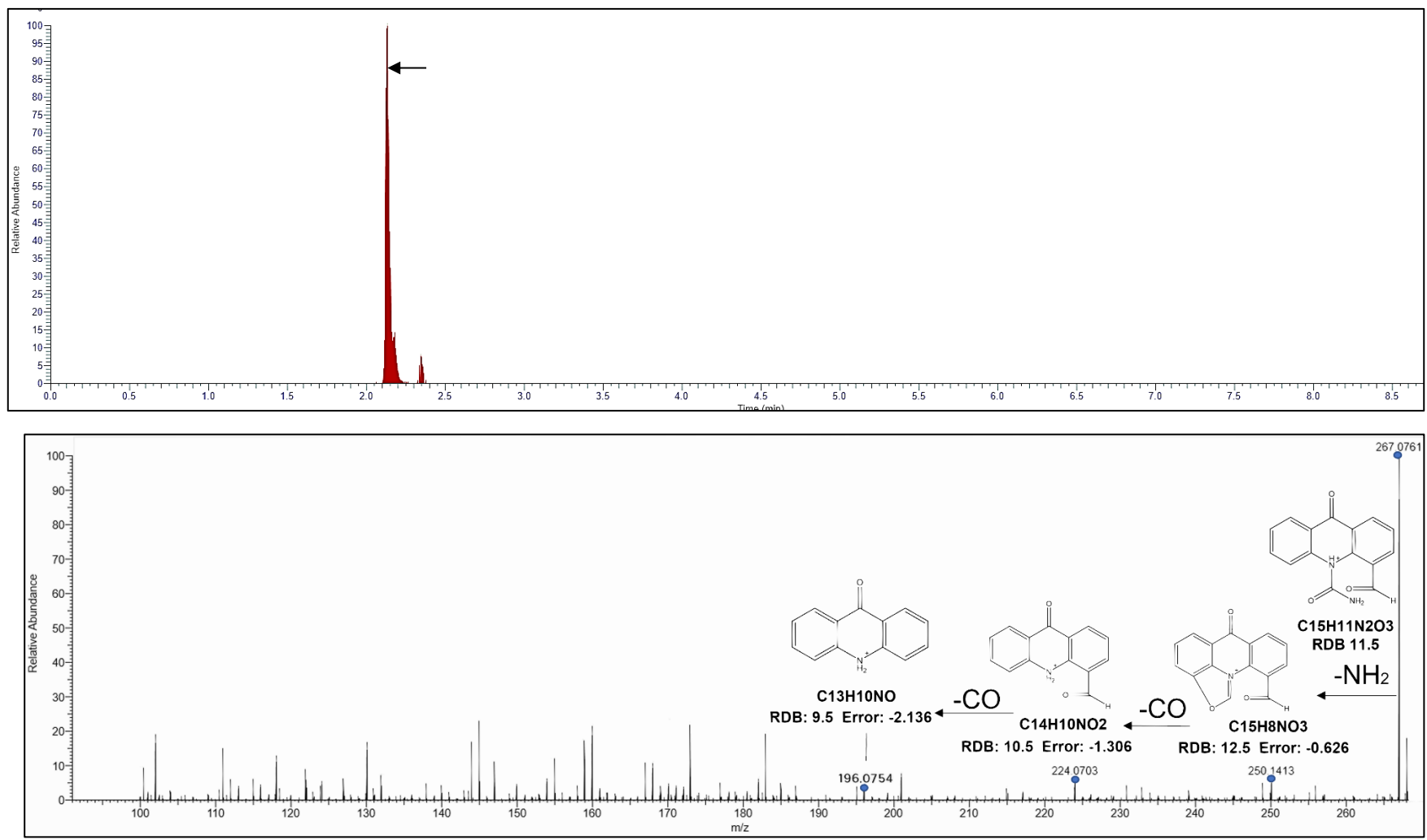

**Figure S13.** MS$^2$ mass spectra of the molecular ion CBZ-266 (a) with proposed structures and losses of fragment ions.

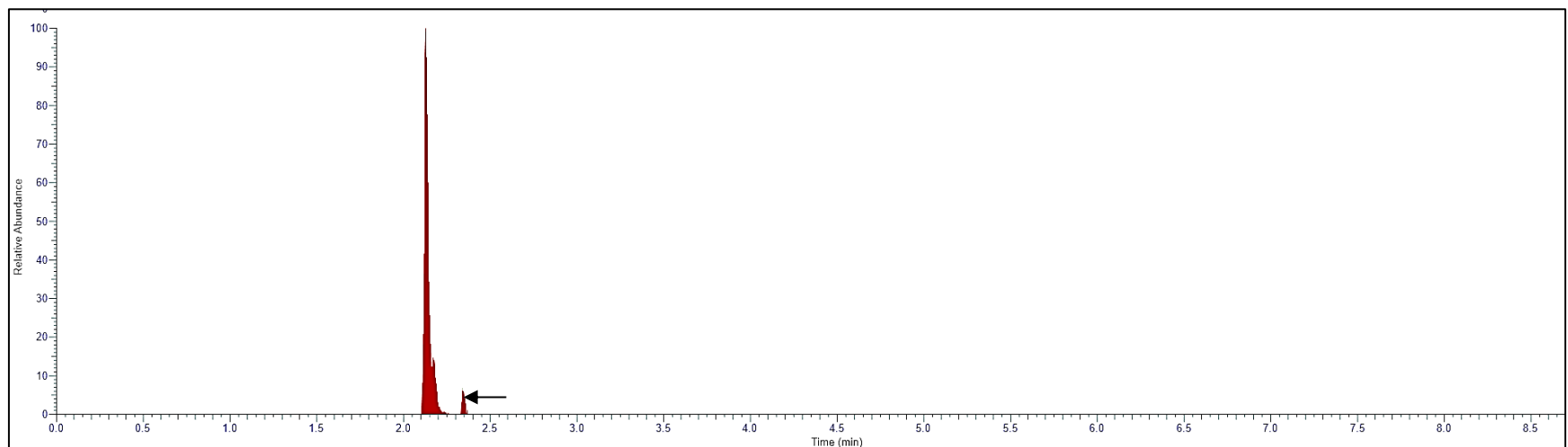

**Figure S14.** MS$^2$ mass spectra of the molecular ion CBZ-266 (b) with proposed structures and losses of fragment ions.

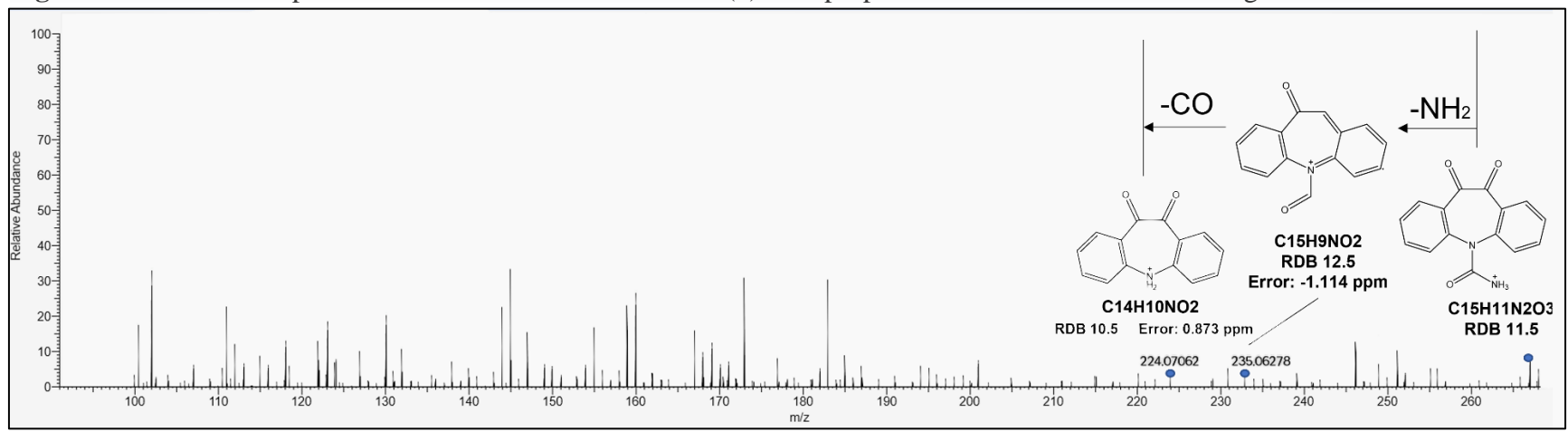

**Table S1.** Chemical structures and physico-chemical properties of the target contaminants: molecular weight (MW), pKa, octanol-water distribution coefficient calculated based on chemical structure at pH 7.4 (ACD/logD), and polar surface area. The information was obtained from https://chemicalize.com/.

| Organic compound MW (g mol$^{-1}$) | Chemical structure | pka | logD | Polar surface area (Å$^2$) |
|---|---|---|---|---|
| **ACY** (225.2) | 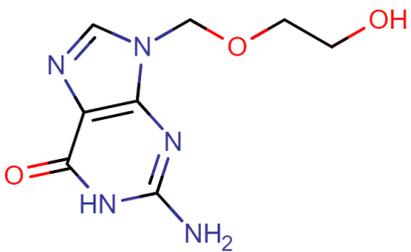 | 2.52; 9.35 | 1.76 | 114.76 |
| **CBZ** (236.3) | 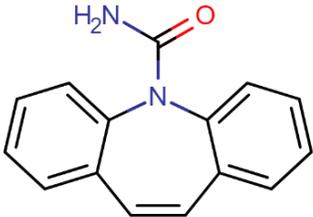 | 15.96 | 2.3 | 46 |
| **DCL** (296.1) | 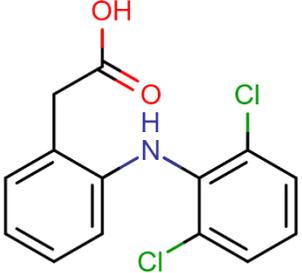 | 4.15 | 1.31 | 49.33 |
| **TCL** (289.5) | 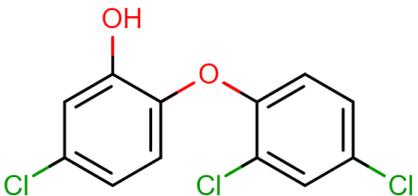 | 7.9 | -1.0 | 59 |
| **TMD** (263.3) | 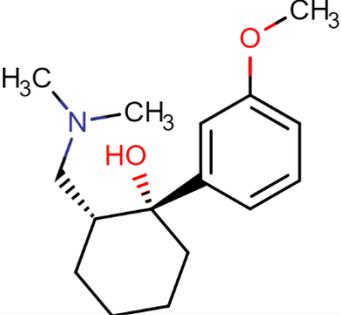 | 9.41 | 1.13 | 32.7 |

**Table S2.** The optimized compound-dependent MS parameters: declustering potential (DP), collision energy (CE) and cell exit potential (CXP) for each compound and each transition of the negative and positive mode.

| Organic compound | Q1 Mass (Da) | Q3 Mass (Da) | DP | CE | CXP |
|---|---|---|---|---|---|
| ACY | 226.10 | 152 | 56 | 19 | 10 |
| ACY | 226.10 | 134.9 | 56 | 41 | 8 |
| CBZ | 237.01 | 194.1 | 156 | 47 | 10 |
| CBZ | 231.01 | 193 | 156 | 47 | 10 |
| DCL | 293.92 | 250 | -65 | -16 | -11 |
| DCL | 293.92 | 214.1 | -65 | -28 | -11 |
| TCL | 287.00 | 35.4 | -40 | -35 | -7 |
| TCL | 289.00 | 37.4 | -40 | -35 | -7 |
| TMD | 264.15 | 58 | 180 | 47 | 10 |
| TMD | 264.15 | 42 | 180 | 47 | 10 |

**Table S3.** Limits of detection (LOD) and quantification (LOQ) of model contaminants at Orbitrap™-high-resolution mass spectrometry (HRMS).

| Organic compound | LOD (µM) | LOQ (µM) |
|---|---|---|
| ACY | 0.49 | 0.14 |
| CBZ | 0.03 | 0.09 |
| DCL | 0.08 | 0.03 |
| TCL | 0.17 | 0.05 |
| TMD | 0.27 | 0.08 |

**Table S4.** Binding Energy (eV), FWHM (eV), Peak area (counts) and Atomic Percentages (%) of O1s XPS spectra obtained for Ti/TiO$_2$ NTA – Mn$_x$O$_y$ and Ti/TiO$_2$ NTA – Mn$_x$O$_y$-Mo electrodes.

| Element | State | Binding energy, eV | | FWHM, eV | | Peak area, counts | | Atomic percentage, % | |
|---|---|---|---|---|---|---|---|---|---|
| | | Mn$_x$O$_y$ | Mn$_x$O$_y$ Mo | Mn$_x$O$_y$ | Mn$_x$O$_y$ Mo | Mn$_x$O$_y$ | Mn$_x$O$_y$ Mo | Mn$_x$O$_y$ | Mn$_x$O$_y$ Mo |
| O1s | Lattice | 529.7 | 529.7 | 1.19 | 1.19 | 7998 | 6025 | 47.3 | 33.9 |
| | Non-lattice | 531.1 | 530.7 | 1.95 | 1.95 | 6150 | 7594 | 36.35 | 42.7 |
| | H$_2$O | 532.6 | 532.3 | 3.28 | 3.20 | 2768 | 4175 | 16.35 | 23.4 |

**Table S5**. The total dissolved manganese measured after open circuit (OC) experiment performed at Ti/TiO$_2$ NTA – Mn$_x$O$_y$ or Ti/TiO$_2$ NTA – Mn$_x$O$_y$-Mo anode using 5 mM NaNO$_3$ electrolyte (pH 8.2, σ 1.3 mS cm$^{-1}$) spiked with contaminants after 2 hours of experiment.

| Material | Mn$^{2+}$, mg L$^{-1}$ |
|---|---|
| Ti/TiO$_2$NTA-Mn$_x$O$_y$ | 5.86±0.05 |
| Ti/TiO$_2$NTA-Mn$_x$O$_y$-Mo | 0.17±0.07 |

**Table S6**. Removal efficiency of ACY, CBZ, DCL, TCL and TRM after 2 hours of experiment performed at 1.2 – 1.4 V/SHE applied to Ti/TiO$_2$ NTA – Mn$_x$O$_y$ or Ti/TiO$_2$ NTA – Mn$_x$O$_y$-Mo anode using 5 mM NaNO$_3$ supporting electrolyte (pH 8.2, σ 1.3 mS cm$^{-1}$).

| Contaminant | Removal efficiency, % | | | |
|---|---|---|---|---|
| | at 1.2 V | | at 1.4 V | |
| | Mn$_x$O$_y$ | Mn$_x$O$_y$-Mo | Mn$_x$O$_y$ | Mn$_x$O$_y$-Mo |
| ACY | 78.88 | 79.24 | 75.81 | 87.10 |
| CBZ | 61.85 | 97.02 | 99.24 | 100.00 |
| DCL | 97.23 | 99.92 | 99.92 | 99.94 |
| TCL | 99.48 | 99.68 | 99.70 | 99.81 |
| TMD | 13.30 | 44.45 | 63.93 | 83.59 |

**Table S7**. The total dissolved manganese measured after complete dissolution of MnO$_2$ coating in ascorbic acid and at 0 - 1.4 V/SHE applied to the Ti/TiO$_2$ NTA – Mn$_x$O$_y$ anode using 5 mM NaNO$_3$ electrolyte spiked with contaminants after 120 min.

| Experiment | Mn$^{2+}$, mg L$^{-1}$ | % of Mn$_x$O$_y$ lost |
|---|---|---|
| Completely dissolved Mn$_x$O$_y$ coating | 36.08±0.6 | |
| after OC | 5.86±0.05 | 16.25 |
| after 1 V | 4.16±2.84 | 11.54 |
| after 1.2 V | 0.47±0.05 | 1.31 |
| after 1.4 V | 0.02±0.01 | 0.34 |

**Table S8**. Observed reaction rates and reaction rates normalized to the electroactive surface area calculated for ACY, CBZ, DCL, TCL and TRM oxidation after 2 hours of experiment performed at 1.2 – 1.4 V/SHE applied to Ti/TiO$_2$ NTA – Mn$_x$O$_y$ or Ti/TiO$_2$ NTA – Mn$_x$O$_y$-Mo anode using 5 mM NaNO$_3$ supporting electrolyte (pH 8.2, σ 1.3 mS cm$^{-1}$).

| Contaminant | at 1.2 V | | | | at 1.4 V | | | |
|---|---|---|---|---|---|---|---|---|
| | Observed reaction rate, h$^{-1}$ | | Normalized reaction rate, m$^3$ h$^{-1}$ m$^{-2}$ | | Observed reaction rate, h$^{-1}$ | | Normalized reaction rate, m$^3$ h$^{-1}$ m$^{-2}$ | |
| | Mn$_x$O$_y$ | Mn$_x$O$_y$-Mo | Mn$_x$O$_y$ | Mn$_x$O$_y$-Mo | Mn$_x$O$_y$ | Mn$_x$O$_y$-Mo | Mn$_x$O$_y$ | Mn$_x$O$_y$-Mo |
| ACY | 0.79 | 0.75 | 3.9 10$^{-4}$ | 2.6 10$^{-4}$ | 0.71 | 1.01 | 3.5 10$^{-4}$ | 3.6 10$^{-4}$ |
| CBZ | 0.49 | 1.73 | 2.5 10$^{-4}$ | 6.71 10$^{-4}$ | 2.36 | 2.42 | 11.8 10$^{-4}$ | 8.6 10$^{-4}$ |
| DCL | 1.44 | 3.32 | 7.2 10$^{-4}$ | 11.8 10$^{-4}$ | 4.03 | 5.10 | 20.1 10$^{-4}$ | 18.2 10$^{-4}$ |
| TCL | 2.82 | 3.23 | 14.1 10$^{-4}$ | 11.3 10$^{-4}$ | 5.09 | 11.51 | 25.4 10$^{-4}$ | 41.1 10$^{-4}$ |
| TMD | 0.06 | 0.28 | 0.3 10$^{-4}$ | 1.09 10$^{-4}$ | 0.49 | 0.86 | 2.4 10$^{-4}$ | 3.1 10$^{-4}$ |

**Table S9.** Removal efficiency and the observed reaction rate of ACY, CBZ, DCL, TCL, TRM and IPM after 2 hours of experiment performed at 1.2 – 1.4 V/SHE applied to Ti/TiO$_2$ NTA – Mn$_x$O$_y$ anode using NaNO$_3$, NaCl or Na$_3$PO$_4$ (pH 8.2, σ 1.3 mS cm$^{-1}$) supporting electrolytes.

| Contaminant | Removal efficiency, % | | | | | | Observed reaction rate, h$^{-1}$ | | | | | |
| --- | --- | --- | --- | --- | --- | --- | --- | --- | --- | --- | --- | --- |
| | at 1.2 V | | | at 1.4 V | | | at 1.2 V | | | at 1.4 V | | |
| | in NaNO$_3$ | in Na$_3$PO$_4$ | in NaCl | in NaNO$_3$ | in Na$_3$PO$_4$ | in NaCl | in NaNO$_3$ | in Na$_3$PO$_4$ | in NaCl | in NaNO$_3$ | in Na$_3$PO$_4$ | in NaCl |
| ACY | 78.9 | 78.5 | 89.9 | 75.8 | 81.2 | 87.6 | 0.79 | 0.82 | 1.12 | 0.71 | 0.80 | 1.00 |
| CBZ | 61.8 | 33.6 | 55.7 | 99.2 | 43.7 | 99.8 | 0.49 | 0.18 | 0.43 | 2.36 | 0.26 | 2.65 |
| DCL | 97.2 | 64.4 | 96.2 | 99.9 | 85.1 | 99.8 | 1.44 | 0.52 | 1.28 | 4.03 | 0.91 | 4.94 |
| TCL | 99.5 | 89.7 | 99.8 | 99.7 | 86.6 | 99.3 | 2.82 | 1.15 | 3.00 | 2.82 | 0.99 | 2.47 |
| TMD | | | | 63.9 | 45.3 | 69.5 | | | | 0.49 | 0.30 | 0.60 |

**Table S10.** The free chlorine (mg L$^{-1}$) measured during the experiments performed at 1.4 V/SHE applied to the Mn$_x$O$_y$-based anode in 5 mM NaCl supporting electrolyte (pH 8.2, σ 1.3 mS cm$^{-1}$).

| Time, min | Free chlorine, mg L$^{-1}$ |
|---|---|
| 0 | 0 |
| 30 | 0±0.03 |
| 60 | 0.1±0.06 |
| 120 | 0.11±0.09 |

**Table S11.** Transformation products (TPs) identified from CBZ during experiment performed at 1.4V/SHE applied to Ti/TiO$_2$ NTA – Mn$_x$O$_y$-Mo anodes performed in 5 mM NaNO$_3$ supporting electrolytes (pH 8.2, σ 1.3 mS cm$^{-1}$).

| Compound | Formula | Annot. DeltaMass [ppm] | Calc. MW | m/z | RT [min] | RDB | Reference Ion | Ref |
|---|---|---|---|---|---|---|---|---|
| CBZ | C$_{15}$H$_{12}$N$_2$O | -1.16 | 236.09 | 237.1 | 2.61 | 11 | [M+H]$^{+1}$ | |
| CBZ-179 | C$_{13}$H$_9$N | -0.84 | 179.07 | 180.0 | 2.3 | 10 | [M+H]$^{+1}$ | [91] |
| CBZ-223 | C$_{14}$H$_9$NO$_2$ | -1.42 | 223.06 | 224.0 | 1.6 | 11 | [M+H]$^{+1}$ | [87], [86], [85], [91] |
| CBZ-266 (a) | C$_{15}$H$_{10}$N$_2$O$_3$ | -0.98 | 266.06 | 267.0 | 2.12 | 12 | [M+H]$^{+1}$ | [86] |
| CBZ-266 (b) | C$_{15}$H$_{10}$N$_2$O$_3$ | -2.88 | 266.06 | 267.0 | 2.34 | 12 | [M+H]$^{+1}$ | [85,86] |

**Text S1: Analysis of target organic contaminants**

ACY, CBZ, TMD were analyzed in electrospray (ESI) positive mode using an Acquity ultraperformance liquid chromatography (UPLC) HSS T3 column (2.1×50 mm, 1.8 µm, Waters) run at 30°C. The eluents employed were acetonitrile with 0.1% formic acid (eluent A), and milli-Q (LC-MS grade) water with 0.1% formic acid (eluent B) at a flow rate of 0.5 mL min$^{-1}$. The gradient was started at 2% of eluent A that was increased to 20% A by 3 min, further increased to 50% A by 6 min and further increased to 95% A by 7 min. It was kept constant for 2.5 min, before returning to the initial condition of 2% A by 9.5 min. The total run time was 11 min. TCL and DCF were analyzed in the ESI negative mode using an Acquity UPLC© BEH C18 column (2.1×50mm, 1.7µm) from Waters run at 30°C. The eluents for ESI negative mode were milli-Q (LC-MS grade) water containing a mixture of acetonitrile and methanol (1:1, v/v) (eluent A) and 1 mM ammonium acetate (eluent B) at a flow rate of 0.6 mL min$^{-1}$. The gradient was started at 5% A, further increased to 100% A by 7 min and then kept constant for 2 min, before returning to the initial conditions of 5 % A by 9 min. The total run time in the ESI negative mode was 11 min.

The target organic contaminants were analyzed in a multiple reaction monitoring (MRM). The settings for the compound-dependent parameters of each transition are summarized in **Table S1**. The source-dependent parameters were as follows: for the positive mode; curtain gas (CUR), 30 V; nitrogen collision gas (CAD), medium; source temperature (TEM), 650°C; ion source gases GS1, 60 V and GS2, 50 V; ion spray voltage, 5500V, and entrance potential (EP), 10V. For the negative mode; curtain gas (CUR), 30V; nitrogen collision gas (CAD), medium; source temperature (TEM), 650°C; ion source gases GS1, 60 V and GS2, 70 V; ion spray voltage, -3500 V, and entrance potential (EP), -10V.

**Text S2: LC-Orbitrap-MS analysis**

LC-MS/MS analysis for transformation products (TPs) of CBZ were acquired with an Orbitrap Exploris 120 mass spectrometer (Thermo Fisher Scientific Inc). The TPs were separated using the column Hypersil GOLDTM (50x2.1 mm, particle size 1.9μ, Thermo fisher). The inject volume of sample was 10 μL. The mobile phase consisted of 0.1% formic acid solution (A) and acetonitrile 0.1% formic acid (B) at 0.4 mL min$^{-1}$. The gradient expressed as the ratio of B was as follows: 0–0.2 min, 2%; a linear increase from 2% to 98%; 0.2–4.75 min, hold at 98% until 6 min, follow by a linear decrease from 98% to 2%; 6–9 min. Orbitrap Exploris 120 equipped was used with an electrospray ionization source (ESI), working in positive ionization mode (3500 V). The samples were analyzed in full scan mode, from 100 to 1000 m/z, with Orbitrap MS (resolution of 30,000) to identify suspect *m/z* and further fragmented at normalized collision energy of 30%. The fragmentation spectra were acquired by Orbitrap working at resolution of 15,000 to obtain structural information of the suspected TPs. All the data was acquired and processed with Compound Discoverer™ 3.0 Software.